\documentclass[11pt,a4paper]{article}
\usepackage[utf8]{inputenc} 
\usepackage[english]{babel}
\usepackage[T1]{fontenc}
\usepackage{graphicx,amssymb,amsmath,wasysym}
\usepackage[bookmarks=false]{hyperref}
\usepackage[nosort]{cite}

\begin{document}
\baselineskip=15.5pt

\thispagestyle{empty}

\begin{flushright}
CFTP/10-009\\
\end{flushright}


\begin{center}

{\LARGE\sc{\bf Constraining Dark Matter Properties with Gamma--Rays
    \\[1ex] 
    from the Galactic Center with {\em Fermi}--LAT}}

\vspace*{9mm}
\setcounter{footnote}{0}
\setcounter{page}{0}
\renewcommand{\thefootnote}{\arabic{footnote}}

\mbox{ {\large \bf Nicolás Bernal$^{1,\,2}$ and Sergio
    Palomares-Ruiz$^2$}}

\vspace*{0.9cm}

{\it $^1$Bethe Center for Theoretical Physics and Physikalisches
  Institut,\\ Universit\"at Bonn, Nu\ss allee 12, D-53115 Bonn,
  Germany}\\
\vskip 0.3cm
{\it $^2$Centro de Física Teórica de Partículas (CFTP), Instituto
  Superior Técnico, \\ 
Universidade Técnica de Lisboa, Avenida Rovisco Pais 1, 1049-001
Lisboa, Portugal} \\  
\vskip5mm
e-mails: {\tt nicolas@th.physik.uni-bonn.de,
  sergio.palomares.ruiz@ist.utl.pt} 
\vskip 5mm

\end{center}

\vspace{1cm}

\begin{abstract}
We study the capabilities of the {\it Fermi}--LAT instrument on board
of the {\it Fermi} mission to constrain particle dark matter
properties, as annihilation cross section, mass and branching ratio
into dominant annihilation channels, with gamma-ray observations from
the galactic center. Besides the prompt gamma-ray flux, we also take
into account the contribution from the electrons/positrons produced in
dark matter annihilations to the gamma-ray signal via inverse Compton
scattering off the interstellar photon background, which turns out to
be crucial in the case of dark matter annihilations into $\mu^+\mu^-$
and $e^+e^-$ pairs.  We study the signal dependence on different
parameters like the region of observation, the density profile, the
assumptions for the dark matter model and the uncertainties in the
propagation model.  We also show the effect of the inclusion of a 20\%
systematic uncertainty in the gamma-ray background.  If {\it
  Fermi}--LAT is able to distinguish a possible dark matter signal
from the large gamma-ray background, we show that for dark matter
masses below $\sim$200~GeV, {\it Fermi}--LAT will likely be able to
determine dark matter properties with good accuracy.

\vspace{3ex}

\vspace{5ex}
\end{abstract}

\newpage

\tableofcontents

\newpage

\section{Introduction}
\label{introduction}

There exist compelling astrophysical and cosmological evidences that
a large fraction of the matter in our Universe is non-luminous and
non-baryonic (see Refs.~\cite{Jungman:1995df, Bergstrom:2000pn, 
  Munoz:2003gx, Bertone:2004pz, Bertone:2010} for reviews).  These
observations, and in particular the precise measurements from the
Cosmic Microwave Background and Large Scale Structure, indicate that it
constitutes $\sim 80\%$ of the total mass content of the
Universe~\cite{Dunkley:2008ie, Tegmark:2006az}.  However, despite the
precision of these measurements, the origin and most of the properties
of the dark matter (DM) particle(s) remain a mystery; little is known
about its mass, spin, couplings and its distribution at small scales.
Nevertheless, DM plays a central role in current structure formation
theories, and its microscopic properties have significant impact on
the spatial distribution of mass, galaxies and clusters.  Thus,
unraveling the nature of DM is of critical importance both from the
particle physics and from the astrophysical perspectives.

Many different particles have been proposed as DM candidates, spanning
a very large range in masses, from light
particles~\cite{Preskill:1982cy, Abbott:1982af, Dine:1982ah, 
  Dodelson:1993je, Boehm:2003hm, Boehm:2003ha, Boehm:2003bt,
  Hooper:2003sh, Boehm:2006mi, Farzan:2009ji} to superheavy candidates
at the Planck scale~\cite{Chang:1996vw, Chung:1998zb, Chung:1998ua,
  Ziaeepour:2000rc, Berezinsky:1997hy, Birkel:1998nx, Sigl:1998vz,
  Blasi:2001hr, Sarkar:2001se} (see, e.g., Refs.~\cite{Bertone:2004pz,
  Bergstrom:2009ib} for a comprehensive list).  Nevertheless, a weakly
interacting massive particle (WIMP), with mass lying from the GeV to
the TeV scale, is one of the most popular candidates for the DM of the
Universe.  WIMPs  can arise in extensions of the Standard Model (SM)
such as supersymmetry (e.g., Ref.~\cite{Jungman:1995df}), little Higgs
(e.g., Ref.~\cite{Birkedal:2006fz}) or extra-dimensions models (e.g.,
Ref.~\cite{Hooper:2007qk}) and are usually thermally produced in the 
early Universe with an annihilation cross section (times relative
velocity) of $\langle\sigma v\rangle \sim 3 \times
10^{-26}$~cm$^3$~s$^{-1}$, which is the standard value that provides the
observed DM relic density.

A variety of techniques has been considered to detect DM.  Among these
are collider experiments to produce DM particles or find evidence for
the presence of particles beyond the SM, direct searches for signals
of nuclear recoil of DM scattering off nuclei in direct detection
experiments, and indirect searches looking for the products of DM
annihilation (or decay), which include antimatter, neutrinos and
photons.  Once this is accomplished and DM has been detected, the next
step would be to use the available information to constrain its
properties.  Different approaches have been proposed to determine the DM
properties by using indirect or direct measurements or their
combination~\cite{Dodelson:2007gd, Bernal:2008zk, Bernal:2008cu,
  Jeltema:2008hf, PalomaresRuiz:2010pn, PalomaresRuiz:2010uu,
  Edsjo:1995zc, Cirelli:2005gh, Mena:2007ty, Agarwalla:2011yy,
  Das:2011yr, Lewin:1995rx, Primack:1988zm, Green:2007rb,
  Bertone:2007xj, Shan:2007vn, Drees:2008bv, Green:2008rd,
  Beltran:2008xg, Shan:2009ym, Strigari:2009zb, Peter:2009ak,
  Chou:2010qt, Shan:2010qv}.  In addition, the information that could
be obtained from collider experiments would also be of fundamental
importance to learn about the nature of DM~\cite{Drees:2000he,
  Polesello:2004qy, Battaglia:2004mp, Allanach:2004xn,
  Weiglein:2004hn, Birkedal:2005jq, Moroi:2005zx, Nojiri:2005ph,
  Baltz:2006fm, Arnowitt:2007nt, Belanger:2008yc, Cho:2007qv,
  Arnowitt:2008bz, Cho:2008tj, Baer:2009bu} and could also be further
constrained when combined with direct and indirect detection
data~\cite{Bernal:2008zk, Baltz:2006fm, Bourjaily:2005ax,
  Altunkaynak:2008ry, Bertone:2010rv}.
 
In this work we study the abilities of the {\it Fermi}--LAT instrument
on board of the {\it Fermi} mission to constrain DM properties, as
annihilation cross section, mass and branching ratio into dominant
annihilation channels, by using the current and future observations of
gamma-rays from the Galactic Center (GC) produced by DM annihilations
(see Refs.~\cite{Kosack:2004ri, Tsuchiya:2004wv, Albert:2005kh,
  Aharonian:2006wh, Collaboration:2009tm} for recent observations of
high-energy gamma-rays from the GC by other experiments).  In general,
disentangling the potential DM signal from the background is the first
task to be addressed.  In addition to the usual approach of searching
for spectral signatures above the expected background, {\it Fermi}--LAT
can also make use of anisotropy studies and might be able to
distinguish the spatial distribution of DM-induced gamma-ray signal
from that of the conventional astrophysical
background~\cite{Ando:2005xg, Ando:2006mt, Ando:2006cr, Cuoco:2006tr,
  Hooper:2007be, Cuoco:2007sh, Zhang:2008rs, SiegalGaskins:2008ge, 
  Taoso:2008qz, Fornasa:2009qh, SiegalGaskins:2009ux, Ando:2009fp,
  Zavala:2009zr, Ibarra:2009nw, Hensley:2009gh, Cuoco:2010jb} (see
also Refs.~\cite{Miniati:2007ke, Keshet:2002sw, Ando:2009nk}).
Throughout this work we assume that a significant understanding of the
large gamma-ray background will be achieved by {\it Fermi}--LAT.

Following Refs.~\cite{Serpico:2008ga, Jeltema:2008hf} (see also
Ref.~\cite{Pieri:2009je}), our default squared region of observation
covers a field of view of $20^\circ \times 20^\circ$ around the GC
(RA=266.46$^\circ$ and Dec=-28.97$^\circ$, corresponding to the
position of the brightest source, as in Ref.~\cite{Vitale:2009hr}) and,
in order to model the relevant gamma-ray foregrounds we use {\it
  Fermi}--LAT observations.  Both, for the simulations of the
background and of the potential signal we use the {\it Fermi} Science
Tools (version \texttt{v9r23p1})~\cite{Fermitools}.  As an improvement
with respect to previous works~\cite{Dodelson:2007gd, Bernal:2008zk,
  Bernal:2008cu, Jeltema:2008hf}, to the commonly considered gamma-ray
prompt contribution, we add the contribution from the electrons and
positrons produced in DM annihilations to the gamma-ray spectrum via
inverse Compton scattering off the ambient photon background.  This
gamma-ray emission turns out to be not crucial in order to reconstruct
DM properties for hadronic channels, but it is very important for DM
annihilations into $\mu^+\mu^-$ and $e^+e^-$ pairs, providing
completely wrong results if not included.  As mentioned above, we do
not include by default any uncertainty in the treatment of the
background or the potential signal.  Hence, in order to study the
effect of different uncertainties and assumptions, we also show the
results for two different observational regions, for two DM density
profiles, for the case when systematical uncertainties in the
gamma-ray background are taken into account (yet, in a simplified
way), for different assumptions about the DM model and when
uncertainties in the propagation model are also considered.

The paper is structured as follows.  In Section~\ref{gammas} we
describe the two main components of the gamma-ray emission from DM
annihilations in the GC by reviewing the relevant formulae and
commenting on the approximations we take.  In Section~\ref{background}
we describe the main gamma-ray foregrounds.  We show the {\it
  Fermi}--LAT sensitivity to DM annihilation in
Section~\ref{sensitivity} and present the {\it Fermi}--LAT prospects 
for constraining DM properties in Section~\ref{results}, where we show
the dependences on several uncertainties and assumptions. Finally, we
draw our conclusions in Section~\ref{conclusions}.

\section{Gamma--Rays from the Galactic Center}
\label{gammas}

The differential intensity of the photon signal (photons per energy
per time per area per solid angle) from a given observational region
in the galactic halo ($\Delta\Omega$) from the annihilation of DM
particles has four main different possible origins: from internal
bremsstrahlung and secondary photons (prompt), from Inverse Compton
Scattering (ICS), from bremsstrahlung and from synchrotron emission,
i.e, 
\begin{equation}
\frac{d\Phi_{\gamma}}{dE_\gamma}(E_{\gamma}, \Delta\Omega) = 
\left(\frac{d\Phi_{\gamma}}{dE_\gamma}\right)_{{\rm prompt}} +
\left(\frac{d\Phi_{\gamma}}{dE_\gamma}\right)_{{\rm ICS}} +
\left(\frac{d\Phi_{\gamma}}{dE_\gamma}\right)_{{\rm bremsstrahlung}} +
\left(\frac{d\Phi_{\gamma}}{dE_\gamma}\right)_{{\rm synchrotron}} ~.
\end{equation}
Bremsstrahlung is due to particle interaction in a medium and should
not be confused with internal bremsstrahlung which is an
electromagnetic radiative emission of an additional photon in the
final state and is part of the prompt gamma-rays.  For the energies of
interest here and for the observational regions we consider, this
contribution is expected to be subdominant with respect to
ICS~\cite{Crocker:2010gy}, so for the sake of simplicity we do not 
discuss it any further.  On the other hand, synchrotron radiation
arises from energetic electrons and positrons traversing the Galactic
magnetic field.  For typical WIMP DM masses, the DM-induced
synchrotron signal lies at radio frequencies.  However, these energies 
are well below the range of interest for an experiment as {\it
  Fermi}--LAT.  Hence, we will also neglect this source of gamma rays
in what follows.  We note, however, that this contribution could also
be used to constrain DM~\cite{Regis:2008ij, Zhang:2008rs,
  Borriello:2008gy, Bertone:2008xr, Nardi:2008ix, Ishiwata:2008qy,
  Bergstrom:2008ag, Bringmann:2009ca, Crocker:2010gy, Linden:2010eu}.

\begin{figure}[t]
\begin{center}
\includegraphics[width=16.7cm]{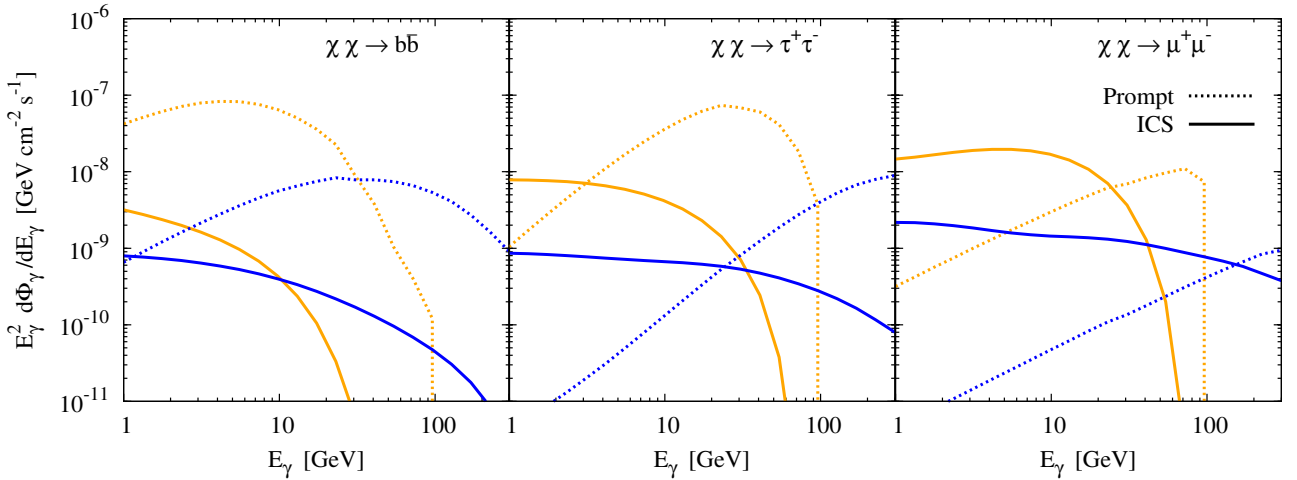}
\end{center}
\caption{\sl \textbf{\textit{Differential signal flux
  $E_\gamma^2\,d\Phi_\gamma/dE_\gamma$ in GeV~cm$^{-2}$~s$^{-1}$ for
  the gamma-ray components}} in the relevant energy range: prompt
  (dotted lines) and ICS (solid lines).  We show the expected results
  for two different DM masses, $m_\chi=105$~GeV (light orange lines)
  and $m_\chi=1$~TeV (dark blue lines).  Each panel depicts a DM
  particle that annihilates into: $b\bar b$ (left panel),
  $\tau^+\tau^-$ (middle panel) and $\mu^+\mu^-$(right panel).  We
  assume a $20^\circ \times 20^\circ$ observational region around the
  GC, a Navarro, Frenk and White DM halo profile, the MED model for
  the propagation model and $\langle\sigma v\rangle=3\cdot
  10^{-26}$~cm$^3$~s$^{-1}$. See the text.}
\label{Fig:spectrum}
\end{figure}

In Fig.~\ref{Fig:spectrum}, we depict the differential signal flux for
$\langle\sigma v\rangle=3\cdot 10^{-26}$~cm$^3$~s$^{-1}$, for three
different annihilation channels ($b\bar b$, $\tau^+\tau^-$ and
$\mu^+\mu^-$) and two different masses, 105~GeV (light orange lines) and
1~TeV (dark blue lines), in a $20^\circ \times 20^\circ$ observational
region around the GC with.  We show the contribution to the signal
from prompt (dotted lines) and ICS (solid lines) gamma-rays.

\subsection{Prompt Gamma--Rays}
\label{prompt}

Whenever DM annihilates into channels with charged particles
in the final states, internal bremsstrahlung photons will unavoidably
be produced. In addition to this, the hadronization, fragmentation,
and subsequent decay of the SM particles in the final states will also
contribute to the total yield of prompt gamma--rays. 

The differential flux of prompt gamma--rays generated from DM
annihilations in the smooth DM halo~\footnote{Throughout this work we
  neglect the contribution due to substructure in the halo, which
  could increase the gamma-ray flux from DM annihilation by a factor of
  $\sim$10~\cite{Diemand:2007qr, Diemand:2008in, Springel:2008cc}.}
and coming from a direction within a solid angle $\Delta\Omega$ can be
written as~\cite{Bergstrom:1997fj}
\begin{equation}
\left(\frac{d\Phi_{\gamma}}{dE_\gamma}\right)_{{\rm prompt}} (E_{\gamma},
\Delta\Omega) = \frac{\langle\sigma
  v\rangle}{2\,m_\chi^2}\sum_i\frac{dN_{\gamma}^i}{dE_{\gamma}}\, 
\text{BR}_i \, \frac{1}{4\,\pi} \, \int_{\Delta\Omega}d\Omega \,
\int_\text{los}\rho\big(r(s,\Omega)\big)^2 \, ds\,, 
\label{Eq:promptflux}
\end{equation}
where the discrete sum is over all DM annihilation channels,
$dN_{\gamma}^i/dE_{\gamma}$ is the differential gamma--ray yield of SM
particles into photons, $\langle\sigma v\rangle$ is the thermal
average of the total annihilation cross section times the relative
velocity, $m_\chi$ is the DM mass, $\rho(r)$ is the DM density
profile, $r$ is the distance from the GC and BR$_i$ is the branching
ratio of DM annihilation into the $i$-th final state.  We simulate the
hadronization, fragmentation and decay of different final states with
the event generator PYTHIA 6.4~\cite{Sjostrand:2006za}, which
automatically includes the so-called final state radiation (photons
radiated off the external legs).  The spatial integration of the
square of the DM density profile is performed along the line
of sight within the solid angle of observation $\Delta\Omega$.  More
precisely, $r = \sqrt{R^2_\odot -  2 s R_\odot \cos \psi + s^2}$,
and the upper limit of integration is $s_{\rm max} = \sqrt{(R_{\rm
    MW}^2 - \sin^2 \psi R^2_\odot)} + R_\odot \cos \psi$, where $\psi$
is the angle between the direction of the galactic center and that of
observation.  As the contributions at large scales are negligible,
different choices of the size of the Milky Way halo, $R_{\rm MW}$,
would not change the results in a significant way.

It is customary to rewrite Eq.~\eqref{Eq:promptflux} introducing the
dimensionless quantity $\overline{J}$, which depends only on the DM
distribution, as 
\begin{equation}
\overline{J}(\Omega) = \frac{1}{\Delta\Omega} \,
\frac{1}{R_\odot\,\rho_\odot^2} \, \int_{\Delta\Omega}d\Omega \,
\int_\text{los} \rho\big(r(s,\Omega)\big)^2\ ds \, ~, 
\label{Eq:Jbarr} 
\end{equation}
where $R_\odot = 8.28$ kpc is the distance from the Sun to the GC
and $\rho_\odot = 0.389$~GeV/cm$^3$ is the local DM
density~\cite{Catena:2009mf}.  The prompt gamma--ray flux can now be
expressed as 
\begin{eqnarray}
\left(\frac{d\Phi_{\gamma}}{dE_\gamma}\right)_{{\rm prompt}} (E_{\gamma},
\Delta\Omega) &=& 4.61 \cdot
10^{-10}\,\mathrm{cm^{-2}\,s^{-1}} \,
\left(\frac{100~\mathrm{GeV}}{m_\chi}\right)^2 \,
\left(\frac{\langle \sigma v\rangle}{3\cdot 10^{-26} {\mathrm{cm^3
      s^{-1}}}}\right) \nonumber \\
 & & \times \left(\frac{{\overline{J}}(\Delta \Omega) \Delta
  \Omega}{\text{sr}}\right) \, \sum_i\frac{dN_{\gamma}^i}{dE_{\gamma}} \,
\text{BR}_i \, ~.
\label{Eq:totpromptflux}
\end{eqnarray}
The value of $\overline{J}(\Delta \Omega) \Delta \Omega$ depends
crucially on the DM distribution.  Detailed structure formation
simulations show that cold DM clusters hierarchically in halos and the 
formation of large scale structure in the Universe can be successfully
reproduced. In the case of spherically symmetric matter density with
isotropic velocity dispersion, the simulated DM profile in the
galaxies can be parameterized via
\begin{equation}
\rho(r) = \rho_\odot \,
\frac{[1+(R_\odot/r_s)^{\alpha}]^{(\beta-\gamma)/\alpha}}
     {(r/R_\odot)^{\gamma}\,[1+(r/r_s)^{\alpha}]^{(\beta-\gamma)/\alpha}}
     ~, 
\label{profile}
\end{equation}
where $r_s$ is the scale radius, $\gamma$ is the inner cusp index,
$\beta$ is the slope as $r \rightarrow \infty$ and $\alpha$ determines
the exact shape of the profile in regions around $r_{\rm s}$.

\begin{table}
\begin{center}
\begin{tabular}{|c|cc|}
\cline{2-3}
\multicolumn{1}{c|}{} &
\multicolumn{2}{c|}{$\overline{J}(\Delta\Omega) \, \Delta\Omega$ [sr]}
\\ 
\multicolumn{1}{c|}{} & $20^\circ \times 20^\circ$ &
$2^\circ \times 2^\circ$ \\[1ex] 
\hline
\hspace{5mm} NFW  \hspace{5mm}   & 11.18 & 1.47  \\[1ex]
\hspace{5mm} Einasto \hspace{5mm} & 19.92 & 2.25  \\ 
\hline
\end{tabular}
\caption{\sl \textbf{\textit{Line of sight integrals of the square of
  the DM density profile.}}  Numerical values of
  $\overline{J}(\Delta\Omega) \, \Delta\Omega$ for the NFW and Einasto
  DM density profiles for two observational regions around the GC:
  $20^\circ \times 20^\circ$ and $2^\circ \times 2^\circ$.
} 
\label{Tab:profiles}
\end{center}
\end{table}

There has been quite some controversy on the values for the $(\alpha,
\beta, \gamma)$ parameters. Commonly used
profiles~\cite{Bahcall:1980fb, Burkert:1995yz, Navarro:1995iw,
  Kravtsov:1997dp, Moore:1999gc} (see also Refs.~\cite{Jaffe:1983iv,
  Evans:1993, Evans:2000, Gentile:2004tb, Salucci:2007tm}) can differ
considerably in the inner part of the galaxy giving rise to important
differences in the final predictions for indirect signals from DM
annihilation. Some N-body simulations suggested highly cusped inner
regions for the galactic halo~\cite{Navarro:1995iw,Moore:1999gc},
whereas others predicted shallower
profiles~\cite{Bahcall:1980fb, Burkert:1995yz, Kravtsov:1997dp}.  The
recent Via Lactea II simulations~\cite{Diemand:2008in} seem to partly
verify earlier results, the so-called Navarro, Frenk and White (NFW)
profile~\cite{Navarro:1995iw} and find their results are well 
reproduced by $(\alpha,\ \beta,\ \gamma)=(1,\ 3,\ 1)$ and $r_s =
20$~kpc.  On the other hand, the Aquarius project simulation
results~\cite{Springel:2008cc} seem to favor a different
parameterization~\cite{Graham:2005xx, Graham:2006ae, 
  Graham:2006af, Navarro:2008kc}, which does not present this effect
of cuspyness towards the center of the Galaxy, the so-called Einasto
profile~\cite{Einasto},
\begin{equation}
 \rho(r) = 0.193 \, \rho_\odot \, \exp\left[ -\frac{2}{\alpha} \left(
   \left(\frac{r}{r_s} \right)^\alpha - 1 \right) \right]\,,\qquad
 \alpha = 0.17 ~, 
\label{Eq:einasto}
\end{equation}
where $r_s = 20$~kpc is a characteristic length.  The values of
$\overline{J}(\Delta \Omega) \Delta \Omega$ for the two observational
regions and for the two DM density profiles under discussion are given
in Table~\ref{Tab:profiles}.

\subsection{Gamma--Rays from Inverse Compton Scattering}
\label{ICS}

Energetic electrons and positrons produced in DM annihilations either
directly or indirectly from the hadronization, fragmentation, and
subsequent decay of the SM particles in the final states, give rise to
secondary photons at various wavelengths via ICS off the ambient
photon background (see Ref.~\cite{Blumenthal:1970gc} for a review).
The differential flux $d\Phi_\gamma/dE_{\gamma}$ of high energy
photons produced by ICS, coming from an angular region of the sky
denoted $\Delta\Omega$, is given by
\begin{equation}
\left(\frac{d\Phi_{\gamma}}{dE_\gamma}\right)_{{\rm ICS}} (E_{\gamma},
\Delta\Omega) = \frac{1}{E_{\gamma}} \, \frac{1}{4\pi} \, 
\int_{\Delta\Omega}d\Omega \, \int_\text{los}ds \,
\int_{m_e}^{m_\chi}dE\,\mathcal{P}(E_{\gamma},E)
\, \frac{dn_e}{dE}\big(E,r_c(s,\Omega),z_c(s,\Omega)\big) ~,
\label{Eq:ICSflux}
\end{equation}
where the differential power emitted into scattered photons of energy
$E_{\gamma}$ by an electron with energy $E$ is given by
$\mathcal{P}(E_{\gamma},E)$ and $dn_e/dE(E,r_c,z_c)$ is the number 
density of electrons and positrons with energy $E$ at a position given
by the cylindrical coordinates $r_c$ and $z_c$, with its origin at the
GC.  The minimal and maximal energies of the electrons are determined
by the electron mass $m_e$ and the DM particle mass $m_\chi$.  The
differential power is defined by
\begin{equation}
\mathcal{P}(E_{\gamma},E) = \frac{3\,\sigma_T}{4\,\gamma^2} \,
E_{\gamma}\,\int_{\frac{1}{4\gamma^2}}^1dq \,
\left[1-\frac{1}{4\,q\,\gamma^2
    \,(1-\tilde\epsilon)}\right] \,
\frac{n_\gamma\left(\epsilon(q)\right)}{q}
\left[2q\,\ln(q)+q+1-2q^2 +
  \frac12\frac{\tilde\epsilon^2}{1-\tilde\epsilon}(1-q)\right] ~,
\label{Eq:P}  
\end{equation}
where $\sigma_T=8\pi\,r_e^2/3\simeq 0.6652$~barn is the total Thomson
cross section in terms of the classical electron radius $r_e$,
$\gamma=E/m_e\gg 1$ is the Lorentz factor of the electron (always
assumed to be relativistic),
$\tilde\epsilon=\frac{E_{\gamma}}{\gamma\,m_e}$ and $\epsilon(q) =  
\frac{m_e}{4\gamma}\frac{\tilde\epsilon}{q\,(1-\tilde\epsilon)}$ is
the energy of the original photon in the system of reference of the
photon gas.

The ambient photon background in Eq.~\eqref{Eq:P}, $n_\gamma(\epsilon)$,
consists of three main components: the cosmic microwave background
(CMB), the starlight concentrated in the galactic plane (SL) and the
infrared radiation due to rescattering of starlight by dust (IR). The
spectrum of the interstellar radiation field (ISRF) in three
dimensions over the whole Galaxy has been calculated in
detail~\cite{Porter:2005qx, Porter:2008ve}.  However, for the sake of
simplicity, in this work we will take an average density field for
each of the regions of observation (but different for each region),
instead of keeping the complete spatial dependence.  Following
Ref.~\cite{Cirelli:2009vg}, we approximate the total radiation density
as a superposition of three blackbody-like spectra,
\begin{equation}
n_\gamma(\epsilon)=\frac{\epsilon^2}{\pi^2} \,
\sum_{i=1}^3\mathcal{N}_i\,\frac{1}{e^{\epsilon/T_i}-1}\, ~,
\end{equation}
with different temperatures and normalizations for each of the three
contributions.  In this work we study two squared regions of observation
around the GC: our default region covers a field of view of $20^\circ
\times 20^\circ$, but we also consider the case of $2^\circ
\times 2^\circ$.  Baring in mind that the observational region of
$20^\circ \times 20^\circ$ around the GC covers most of our default
diffusive zone, we expect the modeling of the ISRF to approximately
provide the correct results for the energy losses that enter in the
diffusion-loss equation (see below).  For instance, this occurs for
the ISRF computed above the galactic plane at $r_c=0$ and
$z_c=5$~kpc~\cite{Porter:2005qx}, which we use to parameterize this
case.  Note that although this position is not located within the
region of observation, it provides very similar results to the ISRF at
$r_c=8$~kpc and $z_c=0$~\cite{Porter:2005qx, Porter:2008ve}.  On the
other hand, for the $2^\circ$-side case, the average photon background
density is expected to be larger, so we model this region with the
computed ISRF on the galactic plane at a distance of 4~kpc from the
GC~\cite{Porter:2008ve}~\footnote{Note, however, that we use the same
  energy losses for all regions of observation as the electrons and
  positrons typically propagate within larger regions.  Thus, to
  compute the effects of propagation, we always take the galactic
  average of the energy losses (see below).}.  For the two regions of
observation under study we use the modelization of the
ISRF~\cite{Porter:2005qx, Porter:2008ve} as given in
Ref.~\cite{Cirelli:2009vg}.  The relevant parameters for the two 
observational regions are given in Table~\ref{Tab:blackbody}.

\begin{table}
\begin{center}
\begin{tabular}{|c|ccc|}
\cline{2-4}
\multicolumn{1}{c|}{} & SL & IR & CMB\\
\hline 
$T_i$           & $0.3$~eV           & $3.5$~meV         & $2.725$ K
\\[1ex] 
$\mathcal{N}_i (20^\circ \times 20^\circ)$ & $8.9\cdot10^{-13}$ &
$1.3\cdot10^{-5}$ &  $1$      \\[1ex]  
$\mathcal{N}_i (2^\circ \times 2^\circ)$ & $2.7\cdot10^{-12}$ &
$7.0\cdot10^{-5}$ & $1$      \\  
\hline 
\end{tabular}
\caption{\sl \textbf{\textit{The parameters of the modelization of the
  ISRF}} taken from Ref.~\cite{Cirelli:2009vg}. The $20^\circ \times
  20^\circ$ and $2^\circ \times 2^\circ$ observational regions around
  the GC are modeled by the ISRF at $(r_c,\,z_c) = (0,\, 5)$~kpc and
  $(r_c,\,z_c) = (4,\, 0)$~kpc, respectively.  See the text.}
\label{Tab:blackbody}
\end{center}
\end{table}

The quantity $dn_e/dE$ in Eq.~\eqref{Eq:ICSflux} is the electron plus
positron spectrum after propagation in the Galaxy (number density per
unit volume and energy), which will differ from the energy spectrum
produced at the source.  We determine this spectrum by solving the
diffusion-loss equation that describes the evolution of the energy
distribution for electrons and positrons assuming steady
state~\cite{Ginzburg:1969}
\begin{equation}
\nabla\Big(K(\vec{x},E)\,\nabla \frac{dn_e}{dE}(\vec{x},E)\Big) +
\frac{\partial}{\partial 
  E}\Big(b(\vec{x},E)\,\frac{dn_e}{dE}(\vec{x},E)\Big) + Q(\vec{x},E)
= 0\,~,  
\label{Eq:masterProp}
\end{equation}
where $K(\vec{x},E)$ is the space diffusion coefficient,
$b(\vec{x},E)$ is the energy loss rate, and $Q(\vec{x},E)$ is the
source term.  The equation above neglects the effect of convection and
reacceleration, which is in general a good approximation for the case
of $e^\pm$~\cite{Delahaye:2008ua}.

The solution of the master equation, Eq.~\eqref{Eq:masterProp}, without
making any simplifying approximations, must be obtained
numerically~\cite{Strong:1998pw}.  However, several assumptions allow 
for semi-analytical solutions of the problem which are able to reproduce
the main features of full numerical approaches and are useful to
systematically study the dependence on the various important
parameters.  Different approaches have been implemented in order to
semi-analytically solve the diffusion equation, in the case that
diffusion and energy losses do not depend on the spatial
coordinates~\cite{Baltz:1998xv, Baltz:2004bb, Colafrancesco:2005ji,
  Lavalle:2006vb, Delahaye:2007fr}.  This is the approach we will
follow. 

The first term in Eq.~\eqref{Eq:masterProp} represents the diffusion of
electrons and positrons and we take the diffusion coefficient as
constant in space, and only depending on energy, $K(E) =
K_0\,\beta\,(E/E_0)^\alpha$, where $K_0$ is the diffusion constant,
$\beta$ is the electron/positron velocity in units of the speed of
light, $\alpha$ is a constant slope, $E$ is the $e^\pm$ energy and
$E_0 = 1$~GeV is a reference energy.

On the other hand, the second term in Eq.~\eqref{Eq:masterProp}
represents the energy losses.  There are different processes that
contribute to these losses: synchrotron radiation, bremsstrahlung,
ionization and ICS.  For electrons and positrons produced in DM
annihilations in the Galaxy, the dominant processes are synchrotron
radiation and ICS.  In the Thomson limit, $b(E) = E^2/(E_0 \tau_E)$,
where $\tau_E = 10^{16}$~s is the characteristic averaged energy-loss
time in the diffusive zone, i.e., energy losses are assumed to have no
spatial dependence. 

Finally, the source term due to DM annihilations in each point of
the halo with DM density profile $\rho(r_c,z_c)$ is given by
\begin{equation}
Q(r_c,z_c,E)=\frac12\,\left(\frac{\rho(r_c,z_c)}{m_\chi}\right)^2
\, \langle\sigma v\rangle \, \sum_i
\text{BR}_i\,\frac{dN_{e^\pm}^i}{dE}\,~,
\end{equation}
where $dN_{e^\pm}^i/dE$ is the prompt electron plus positron spectrum
produced in DM annihilations into channel $i$.

In this work we shall use the popular two-zone diffusion model and
obtain the semi-analytical solution using the Bessel approach as
described in detail in Ref.~\cite{Delahaye:2007fr}.  In this model,
electron and positron propagation takes place in a cylindrical region
(the diffusive zone) around the galactic center of half thickness $L$
and radius $R_\text{gal}$; the propagating particles being free to
escape the region, a case in which they are simply lost.  Regarding 
the propagation parameters $L$, $K_0$ and $\alpha$, we take their
values from the commonly used MIN, MAX and MED
models~\cite{Delahaye:2007fr} (see Table~\ref{Tab:PropParametersPos}),
which correspond to the minimal, maximal and median primary positron
fluxes over some energy range that are compatible with the B/C
data~\cite{Maurin:2001sj}.  However, it has been pointed
out~\cite{Delahaye:2007fr} that the propagation configurations
selected by the B/C analysis do not play the same role for primary 
antiprotons and positrons.  In particular, the MIN configuration for
antiprotons~\cite{Donato:2003xg} does not have an equivalent for
positrons, for which it is not possible to single out one combination
of the parameters which would lead to the minimal value of the
positron signal.  In any case, we consider these three (approximate)
limiting models as a reference to set the uncertainty in the
propagation parameters.

\begin{table}
\begin{center}
\begin{tabular}{|c|ccc|}
\cline{2-4}
\multicolumn{1}{c|}{} &$L$ [kpc]&$K_0$ [kpc$^2$/Myr]&$\alpha$\\
\hline 
MIN & $1$  & $0.00595$ & $0.55$ \\[1ex]  
MED & $4$  & $0.0112$  & $0.70$ \\[1ex] 
MAX & $15$ & $0.0765$  & $0.46$ \\
\hline 
\end{tabular}
\caption{\sl \textbf{\textit{Values of the propagation parameters}}
  that roughly provide minimal, median and maximal $e^\pm$ fluxes
  compatible to the B/C data.}
\label{Tab:PropParametersPos}
\end{center}
\end{table}

The resulting $e^\pm$ flux from DM annihilations can be written
as~\cite{Baltz:1998xv, Delahaye:2007fr}
\begin{equation}
\frac{dn_{e^\pm}}{dE}
(r_c,z_c,E) = \frac{\beta \, \langle\sigma v\rangle}{2}
\left(\frac{\rho(r_c,z_c)}{m_\chi}\right)^2 
\frac{E_0\,\tau_E}{E^2} \, \sum_i \text{BR}_i \, \int_E^{m_\chi}
\frac{dN_{e^\pm}^i}{dE_s} (E_s)
\,\tilde{I}(\lambda_D,r_c,z_c)\,dE_s\,~,  
\label{Eq:PosFlux}
\end{equation}
where the final energy and that at the source are denoted by $E$ and
$E_s$, respectively.  The so-called halo function,
$\tilde{I}(\lambda_D,r_c,z_c)$, contains all the dependence on the
astrophysical factors and is independent on the particle physics
model.  It is given by~\cite{Delahaye:2007fr}
\begin{equation}
\tilde{I}(\lambda_D,r_c,z_c)=\sum_{i,\,n=1}^\infty
J_0\left(\frac{\alpha_i\,r_c}{R_\text{gal}}\right)\,\varphi_n(z_c)\,\exp
\left[-\left\{\left(\frac{n\,\pi}{2L}\right)^2+\frac{\alpha_i^2}
  {R_\text{gal}^2}\right\}\frac{\lambda_D^2}{4}\right]\,R_{i,n}(r_c,z_c)
\,~,  
\end{equation}
where $\lambda_D$ is the diffusion length, defined by 
\begin{equation}
\lambda_D^2(E,\,E_s) = 4\,K_0\,\tau_E
\left(\frac{(E/E_0)^{\alpha-1} - (E_s/E_0)^{\alpha-1}}{1-\alpha}
\right) ~,
\end{equation}
$\alpha_i$'s are the zeros of the Bessel function $J_0$ and
$\varphi_n(z_c)=(-1)^m\,\cos\big(n\,\pi\,z_c/(2L)\big)$ with odd 
$n=2m+1$, which ensures that the halo function vanishes at the
boundaries $z_c=\pm L$.  The coefficients $R_{i,n}$ are the Bessel and
Fourier transforms of the DM density squared:
\begin{equation}
R_{i,n}(r_c,z_c)=\frac{2}{L\,R_\text{gal}^2}\,\frac{1}{J_1^2(\alpha_i)}\,
\int_{-L}^{+L}dz\,\int_0^{R_\text{gal}} r\, dr \,
\left[\frac{\rho(r,z)}{\rho(r_c,z_c)}\right]^2 \, J_0
\left(\frac{\alpha_i\,r}{R_\text{gal}}\right) \, \varphi_n(z)\,~. 
\end{equation}
The advantage of this method is that (for each density profile and
propagation model) the halo function $\tilde{I}(\lambda_D,r_c,z_c)$
can be calculated and tabulated just once as a function of the diffusion
length and the position, and then be easily used for performing
parameter space scans which, as in our case, can be rather large.

\section{Gamma-Ray Foregrounds}
\label{background}

\begin{figure}[t]
\begin{center}
\includegraphics[width=16cm]{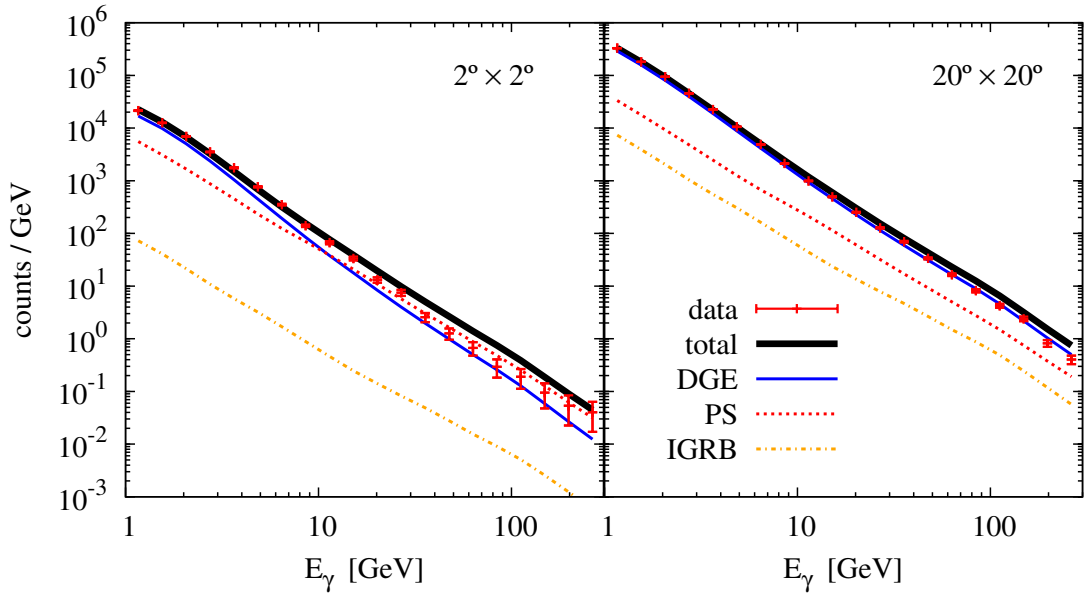}
\end{center}
\caption{\sl \textbf{\textit{Energy spectrum of the gamma-ray
  background.}}  The number of counts observed by {\it Fermi}--LAT from
  August 4, 2008 to July 12, 2011 are shown in red with error bars.
  The three components of the background obtained after performing an
  unbinned likelihood analysis of the data with the {\it Fermi}
  Science Tools (see text) are also depicted: diffuse galactic
  emission (thin solid blue line), isotropic gamma-ray background
  (dashed-dotted orange line), resolved point sources (dashed red
  line) and the total fitted contribution (thick solid black line).
  We show the results for two squared windows around the GC: $2^\circ
  \times 2^\circ$ (left panel) and $20^\circ \times 20^\circ$ (right
  panel).}
\label{Fig:bkg}
\end{figure}

In Fig.~\ref{Fig:bkg} we show the {\it Fermi}--LAT data
obtained from August 4, 2008 (15:43:37 UTC) to July 12, 2011 (09:45:27
UTC) in the two windows around the GC (RA=266.46$^\circ$ and
Dec=-28.97$^\circ$) considered here.  We extract the data from the
{\it Fermi} Science Support Center archive~\cite{Fermidata} and select
only events classified as \texttt{DIFFUSE}, which are the appropriate
ones to perform this analysis with the \texttt{gtselect} tool.  We use a
zenith angle cut of 105$^\circ$ to avoid contamination by the Earth's
albedo and the instrument response function \texttt{P6V11}. There are
three components contributing to the high-energy gamma-ray background:
the diffuse galactic emission (DGE), the isotropic gamma-ray background
(IGRB) and the contribution from resolved point sources (PS).  By
making use of the public {\it Fermi} Science Tools (version
\texttt{v9r23p1})~\cite{Fermitools} we perform an unbinned likelihood
analysis of the data with the \texttt{gtlike} tool and show these
contributions.  In the $2^\circ \times 2^\circ$ region around the GC
(left panel), the background coming from the resolved point sources is
the dominant one for energies above $\sim$~10~GeV. However, for a
$20^\circ \times 20^\circ$ region around the GC (right panel), the DGE
is the most important one.  The IGRB contribution, being at the
percent level or smaller, does not have any effect on the results.
Nevertheless, we have included it.

\subsection{Diffuse Galactic Emission}
\label{EDG}

The DGE is mainly produced by the interactions of cosmic ray nucleons
and electrons with the interstellar gas, via the decay of neutral
pions and bremsstrahlung, respectively, and by the inverse Compton
scattering of cosmic ray electrons with the ISRF.  We also note that the
contribution from unresolved point sources is expected to be
small~\cite{Strong:2006hf} and is not taken into account
here.  Assuming that the cosmic ray spectra in the Galaxy can be
normalized to the solar system measurements, the so-called conventional
model was derived~\cite{Strong:1998pw, Moskalenko:1997gh,
  Strong:2004de}.  This model failed to reproduce the measurements by
the EGRET experiment, in particular in the GeV range where the data
shows an excess~\cite{Hunger:1997we}.  However, recent measurements by
{\it Fermi}--LAT show no excess and are well reproduced by the
conventional model, at least at intermediate galactic latitudes 
$10^\circ < |b| < 20^\circ$ and up to 10~GeV~\cite{Abdo:2009ka,
  Abdo:2009mr}.  In order to model this foreground we use the {\it
  Fermi}-LAT model map
\texttt{gll\_iem\_v02\_P6\_V11\_DIFFUSE.fit}~\cite{Fermimodels}.

\subsection{Isotropic Gamma-Ray Background}
\label{IGRB}

A much fainter and isotropic diffuse emission was first detected by
the $SAS$--$2$ satellite~\cite{Fichtel78} and later confirmed by the
measurement reported by the EGRET experiment~\cite{Sreekumar:1997un}.
Although the term extragalactic gamma-ray background is commonly used,
the extragalactic origin for this component is not clearly
established~\cite{Keshet:2003xc, Moskalenko:2006ta, Orlando:2008uk, 
  Moskalenko:2009tv}, so we use the term isotropic gamma-ray
background (IGRB).  Among the possible contributions to this emission, 
we can list unresolved extragalactic sources such as blazars, active
galactic nuclei, starbursts galaxies, star forming galaxies, galaxy
clusters, clusters shocks and gamma-ray bursts~\cite{Stecker:1996ma,
  Pavlidou:2002va, Gabici:2002fg, Totani:1998xc} as well as other
processes giving rise to truly diffuse emission~\cite{Loeb:2000na,
  Kalashev:2007sn}.

Recently the {\it Fermi}--LAT collaboration reported a new measurement
of this high-energy gamma-ray emission~\cite{Abdo:2010nz}, consistent
with a power law with differential spectral index $\alpha=2.41\pm
0.05$ and intensity $\Phi(E_\gamma>100$ MeV$)=(1.03\pm0.17)\cdot 10^{-5}$
cm$^{-2}$ s$^{-1}$ sr$^{-1}$~\cite{Abdo:2010nz}, i.e., for the
best-fit values, 
\begin{equation}
\left(\frac{d\Phi}{dE_\gamma}\right)_\text{IGRB} (E_\gamma) = 5.65 \cdot
10^{-7}\cdot
\left(\frac{E_\gamma}{\text{GeV}}\right)^{-2.41}\quad\text{GeV}^{-1}\, 
\text{cm}^{-2}\,\text{s}^{-1}\,\text{sr}^{-1}\,. 
\end{equation}
Let us note, however, that this spectrum is significantly softer than
the measured EGRET spectrum with index $\alpha_\text{EGRET}=2.13\pm
0.03$~\cite{Sreekumar:1997un}.

To simulate this background, we use the model map
\texttt{isotropic\_iem\_v02\_P6\_V11\_DIFFUSE.txt} supplied by the
collaboration~\cite{Fermimodels}.

\subsection{Point Sources}
\label{PS}

Finally, another important source of background particularly important
when looking at the GC is that of resolved point sources.  We consider
the catalog of high-energy gamma-ray sources obtained by the first
11~months of {\it Fermi}--LAT (from August 2008 to July 2009).  It 
contains 1451 sources detected with a significance better than
$4\sigma$ in the $100$~MeV to $100$~GeV
range~\cite{Collaboration:2010ru} and it represents a considerable 
increase with respect to the third EGRET catalog~\cite{Hartman:1999fc}
which contained 271 sources.  All spectra have been fitted by power
laws and we explicitly include all these sources in our analysis.
Ideally, the total emission corresponds to a superposition of all the
spectra within the region of observation
\begin{equation}
\left(\frac{d\Phi}{dE_\gamma}\right)_\text{PS} (E_\gamma,l,b) =
\sum_{i\,\in \Delta\Omega}
\phi_i\,\left(\frac{E_\gamma}{\text{GeV}}\right)^{-\alpha_i} ~. 
\end{equation}
However, due to the finite angular resolution of the experiment, in
order to perform the fit, we have also included the contribution from
point sources outside the region of observation.  Note also that the
{\it Fermi}--LAT catalog includes some sources which are found in
regions with bright or possibly incorrectly modeled diffuse emission
which could affect the measured properties of these sources, as well
as sources with some inconsistency.  Conservatively, we have also
included these sources in our analysis.

\section{{\em Fermi}--LAT sensitivity to DM annihilation}
\label{sensitivity}

The {\it Fermi Gamma-ray Space Telescope (Fermi)}~\cite{Atwood:2009ez}
was launched in June $2008$ for a mission of 5~to 10~years.  The Large
Area Telescope ({\it Fermi}--LAT) is the primary instrument on board of
the {\it Fermi} mission. It performs an all-sky survey, covering a
large energy range for gamma-rays (from below 20~MeV to more than
$300$~GeV), with an energy and angle-dependent effective area which to
good approximation is $A_\text{eff} \simeq 8000$~cm$^2$ and a field of
view $\text{FoV}=2.4$~sr.  Its equivalent Gaussian $1\sigma$ energy
resolution is $\sim 10\%$ at energies above 1~GeV.  In order to model
the signal, we make use of the public {\it Fermi} Science Tools (version
\texttt{v9r23p1})~\cite{Fermitools}, which allow us to properly
simulate the performance of the experiment, taking into account the
energy resolution, the point spread function, the dependence of the
effective area with the energy, etc.  In the following analysis, we
consider a 5-year mission run, and an energy range from 1~GeV
extending up to 300~GeV, divided with \texttt{gtbin} into 20 evenly
spaced logarithmic bins.  We generate photon events from DM
annihilations according to the instrument response function by means
of \texttt{gtobssim}.

On the other hand, the optimal size of the region of observation
around the GC depends on different factors, from pure geometrical ones
to the presence of the different type of foregrounds with different
spatial dependences.  It has been pointed out that in order to
maximize the signal-to-noise ratio, the best strategy is to focus on a
(squared) region around the GC with a $\sim10^\circ$ side for a NFW
profile~\cite{Serpico:2008ga, Jeltema:2008hf}.  Hence, we choose a
squared region with a $10^\circ$ side around the GC ($20^\circ \times
20^\circ$) as our default region of observation.  However, we will
also illustrate some results for the case $2^\circ \times 2^\circ$.

\begin{figure}[t]
\begin{center}
\includegraphics[width=16.7cm]{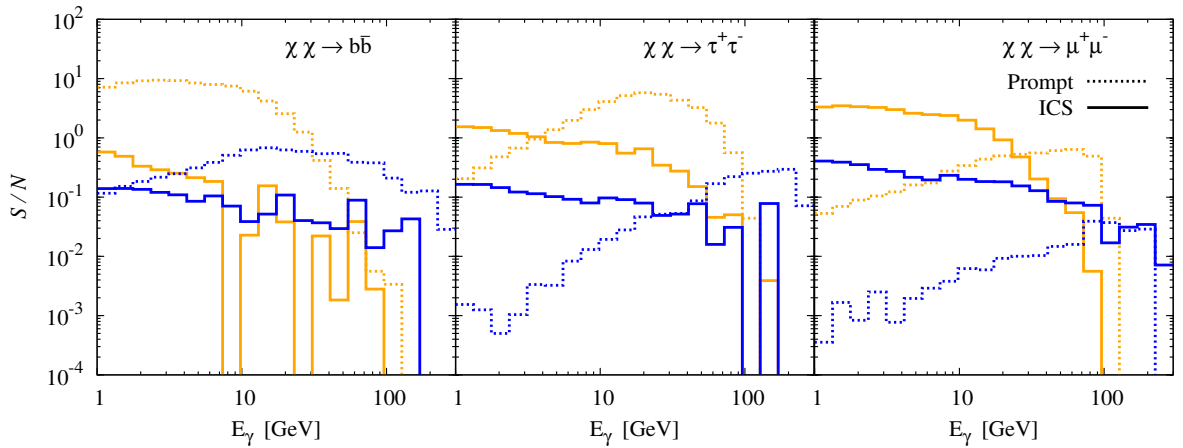}
\end{center}
\caption{\sl \textbf{\textit{The signal-to-noise ratio for all energy
  bins.}}  We show in each panel the case of two DM masses,
  $m_\chi=105$~GeV (light orange lines) and $m_\chi=1$~TeV (dark blue
  lines), and the individual results from the prompt (dotted lines)
  and ICS (solid lines) emission.  The three panels correspond to DM
  annihilation into $b\bar b$ (left panel), $\tau^+\tau^-$ (middle
  panel) and $\mu^+\mu^-$ (right panel).  We assume 5~years of data
  taking with {\it Fermi}--LAT, a $20^\circ \times 20^\circ$
  observational region around the GC, a NFW DM halo profile, the MED
  model for the propagation model and $\langle\sigma v\rangle=3\cdot
  10^{-26}$~cm$^3$~s$^{-1}$.} 
\label{Fig:signaltonoise}
\end{figure}

By looking at the signal-to-noise ratio, $S/N$, where $S$ is the
signal and $N=\sqrt{B}$, with $B$ being the background, one can also
easily understand the relevance of the different components of the
signal for each given set of the values of the parameters.  We show
the ratio $S/N$ in Fig.~\ref{Fig:signaltonoise} for three different
annihilation channels ($b\bar b$, $\tau^+\tau^-$ and $\mu^+\mu^-$) and
two different masses (105~GeV and 1~TeV) after 5~years of data taking
in a $20^\circ \times 20^\circ$ region around the GC.  We show the
contribution to this ratio from prompt and ICS gamma-rays.  As it is
evident from the figure, for the case of DM annihilation into $b\bar
b$ (or more generically, into hadronic channels), the contribution
from ICS gamma-rays is always subdominant with respect to that from
prompt gamma-rays.  This was already apparent from
Fig.~\ref{Fig:spectrum}.  However, from Fig.~\ref{Fig:spectrum} one
would na\"{\i}vely expect that for the case of DM annihilation into
leptonic channels, the inclusion of the ICS contribution would have a
very important effect in the results (the heavier the DM the more
important), as the total yields come mainly from this component.  Yet,
this is not what Fig.~\ref{Fig:signaltonoise} shows for the
$\tau^+\tau^-$ channel, whose ICS contribution is not as important as
that from the $\mu^+\mu^-$ (and, not shown here, the $e^+e^-$)
channel.  This is easy to understand by recalling
Figs.~\ref{Fig:spectrum} and~\ref{Fig:bkg}.  For the energy window of
observation and the parameters considered, the gamma-ray signal is
dominated at low energies by ICS, but it is also at low energies where
the background is higher.  On the other hand, the background drops
faster with energy than the signal, so even with fewer counts in the
detector, the $S/N$ ratio is larger at high energies where the prompt
contribution is more important.  The case of the $\mu^+\mu^-$ (and
$e^+e^-$) channel is different in this regard; the prompt spectrum is
harder than in the $\tau^+\tau^-$ case and the yields are lower.  As a
consequence the ICS contribution dominates the signal up to much
higher energies and hence its inclusion in the analysis renders
necessary.

In order to evaluate the sensitivity of {\it Fermi}--LAT to DM
annihilation we perform an analysis in terms of a $\chi^2$ function,
defined by~\footnote{In principle, this expression is only valid if,
  at least, there are several photons in each energy bin (the Gaussian
  limit).  Although a priori this is not guaranteed, this is the
  actual situation for all the models above the sensitivity curves.}
\begin{equation}
\chi^2 \left(m_\chi, \, \langle\sigma v\rangle\right) =
\sum_{i=1}^{20}\frac{\big(S_i \left(m_\chi, \, \langle\sigma
  v\rangle\right)\big)^2}{B_i} ~, 
\end{equation}
where $S_i$ the number of signal (DM-induced) events in the $i$-th
energy bin and $B_i$ the corresponding background events in the same
bin.  Here we are assuming perfect knowledge of the background, but as
mentioned in Ref.~\cite{Baltz:2008wd}, an addition of a 20\%
uncertainty in the modeled background would only worsen the
sensitivity by about 5\%.  However, let us note that the results
presented here represent the most optimistic case.  If the background
is simultaneously fitted with the signal the results would worsen by an
${\cal O}(1)$ factor.

\begin{figure}[t]
\begin{center}
\includegraphics[width=16.7cm]{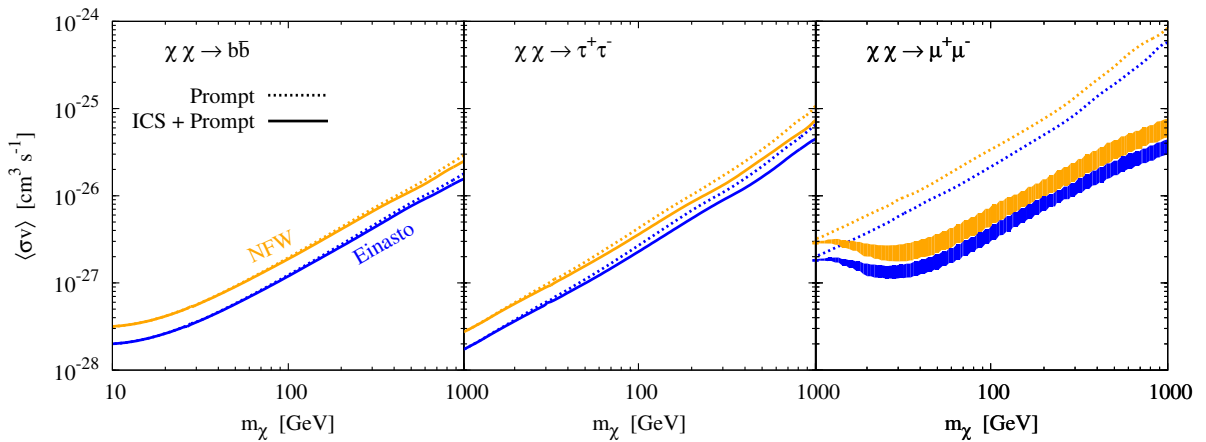}
\end{center}
\caption{\sl \textbf{\textit{ Fermi-LAT sensitivity.}} The regions on
  the parameter space $(m_\chi,\,\langle\sigma v\rangle)$ that could
  be probed by the gamma ray measurements by {\it Fermi}--LAT after
  5~years of data taking in a $20^\circ \times 20^\circ$ observational
  region around the GC.  The three panels correspond to DM
  annihilation into $b\bar b$ (left panel), $\tau^+\tau^-$ (middle
  panel) and $\mu^+\mu^-$ (right panel).  The solid (dotted) lines
  correspond to the $90\%$~CL contours (1 dof) with (without) the ICS
  contribution and the light orange (dark blue) lines correspond to a
  NFW (Einasto) DM halo profile.}
\label{Fig:exclu}
\end{figure}

In Fig.~\ref{Fig:exclu} we depict the sensitivity plots for a
$20^\circ \times 20^\circ$ observational region around the GC in
the plane $(m_\chi, \,\langle\sigma v\rangle)$ for the three
annihilation modes: $b\bar b$, $\tau^+\tau^-$ and $\mu^+\mu^-$.  The
solid (dotted) lines correspond to the $90\%$ confidence level
(CL) for 1 degree of freedom (dof) contours with (without) the ICS
contribution and the light orange (dark blue) lines correspond to a
NFW (Einasto) DM halo profile.  The solid bands represent the
uncertainty in the propagation parameters, which are more important
for the $\mu^+\mu^-$ case for which the contribution from ICS clearly
affects the results, improving the expectations of detection by an
order of magnitude at high masses.  Note however, that the uncertainty
in the propagation parameters is smaller~\footnote{For a $2^\circ
  \times 2^\circ$ observational region around the GC, these
  uncertainties are more important and make the bands slightly wider.}
than what would na\"{\i}vely be expected from the differences in the
halo function for the three propagation models (see e.g.,
Ref.~\cite{Delahaye:2007fr}).  Our findings agree with other related
results, although obtained for a different observational region than
the one considered in this study~\cite{Cumberbatch:2010ii}.  The
regions above these lines represent the sets of parameters that could
be probed by {\it Fermi}--LAT after 5~years of data taking.  Notice that
Ref.~\cite{Baltz:2008wd} (see also Ref.~\cite{Dodelson:2007gd}) 
performed a similar analysis considering a region of $0.5^\circ$
around the GC, which for a NFW density profile is expected to provide
worse results by a factor of $\sim$~7--8~\cite{Jeltema:2008hf,
  Serpico:2008ga}.  In addition to the different confidence level
considered, there are also small differences with our assumptions in the
binning, density profile parameters, etc.  Finally, let us note that
there is a minimum in the sensitivity curve when ICS is included in
the $\mu^+\mu^-$ case.  This can be understood by the fact that below
this mass, the signal-to-noise ratio due to the prompt signal starts
to become comparable in the range of energies considered in this
analysis (1~GeV--300~GeV) and thus, this curve tends to the case with
only prompt gamma-rays.

As a general remark, the hadronic channels generate a higher yield of
photons, giving rise to the most optimistic results.  As already
mentioned, in this case, the addition of the ICS contribution does not 
improve the results because it is always subdominant, even for large
values of the DM mass (left panel of Fig.~\ref{Fig:exclu}).  On the
other hand, the inclusion of the ICS contribution renders very
important in the case of DM annihilations into the $\mu^+\mu^-$ (and,
not shown here, the $e^+e^-$) channel.

\begin{figure}[t]
\begin{center}
\includegraphics[width=8.1cm]{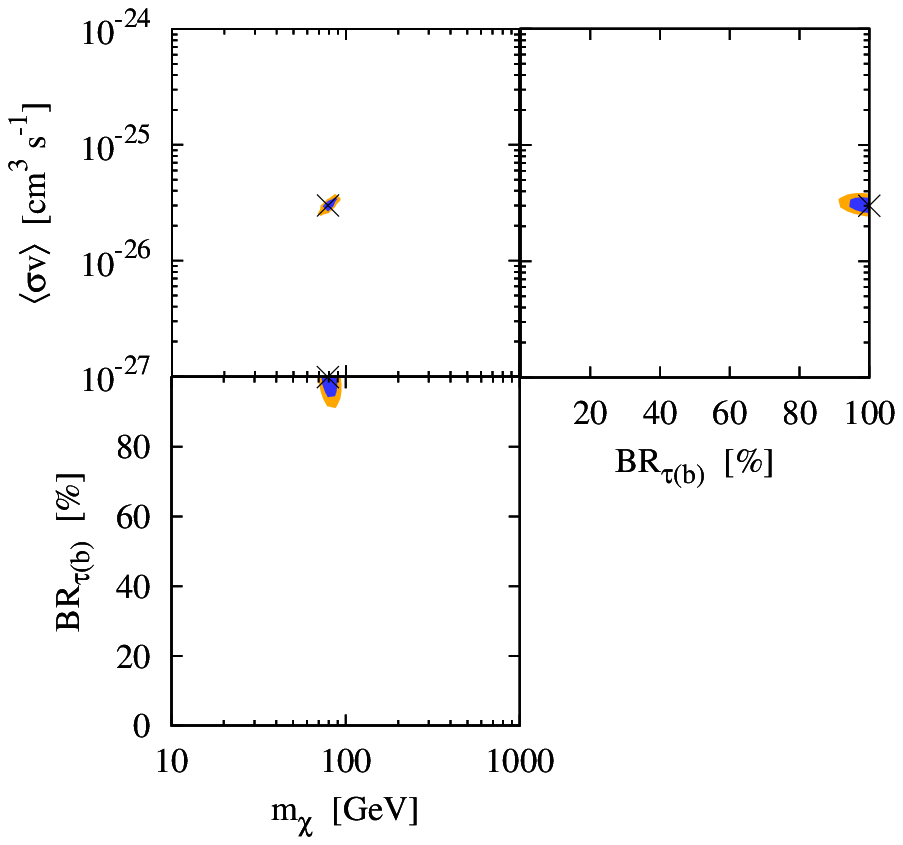}
\includegraphics[width=8.1cm]{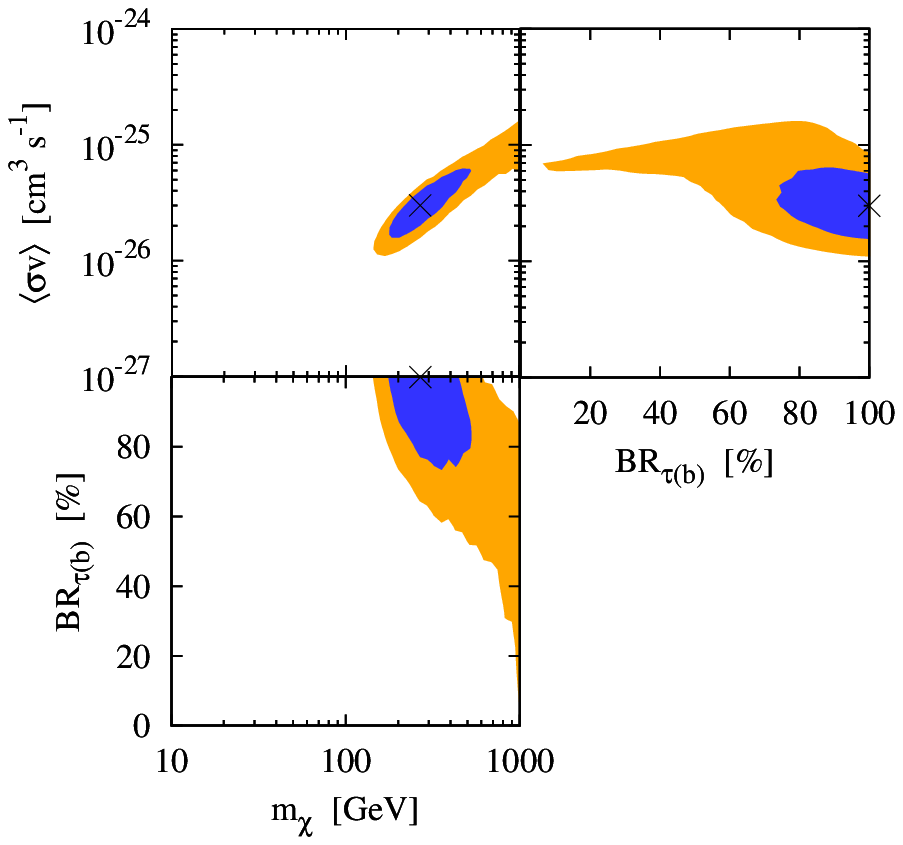}
\end{center}
\caption{\sl \textbf{\textit{Fermi--LAT abilities to constrain DM
  properties.}}  We consider DM annihilation into a pure
  $\tau^+\tau^-$ final state and two DM masses: $m_\chi^0=80$~GeV
  (left panels) and $m_\chi^0=270$~GeV (right panels).  Dark blue
  (light orange) regions represent $68\%$~CL ($90\%$~CL) contours for
  2 dof.  We assume a $20^\circ \times 20^\circ$ observational region
  around the GC, a NFW DM halo profile, the MED propagation model and
  $\langle\sigma v\rangle^0=3\cdot 10^{-26}$~cm$^3$~s$^{-1}$.  The
  parameters in this plot represent our default setup (see
  Table~\ref{Tab:figs}).  The black crosses indicate the values of the
  parameters for the simulated observed ``data''.} 
\label{Fig:tau10deg}
\end{figure}

\section{Constraining DM properties}
\label{results}

\begin{figure}[t]
\begin{center}
\includegraphics[width=8.1cm]{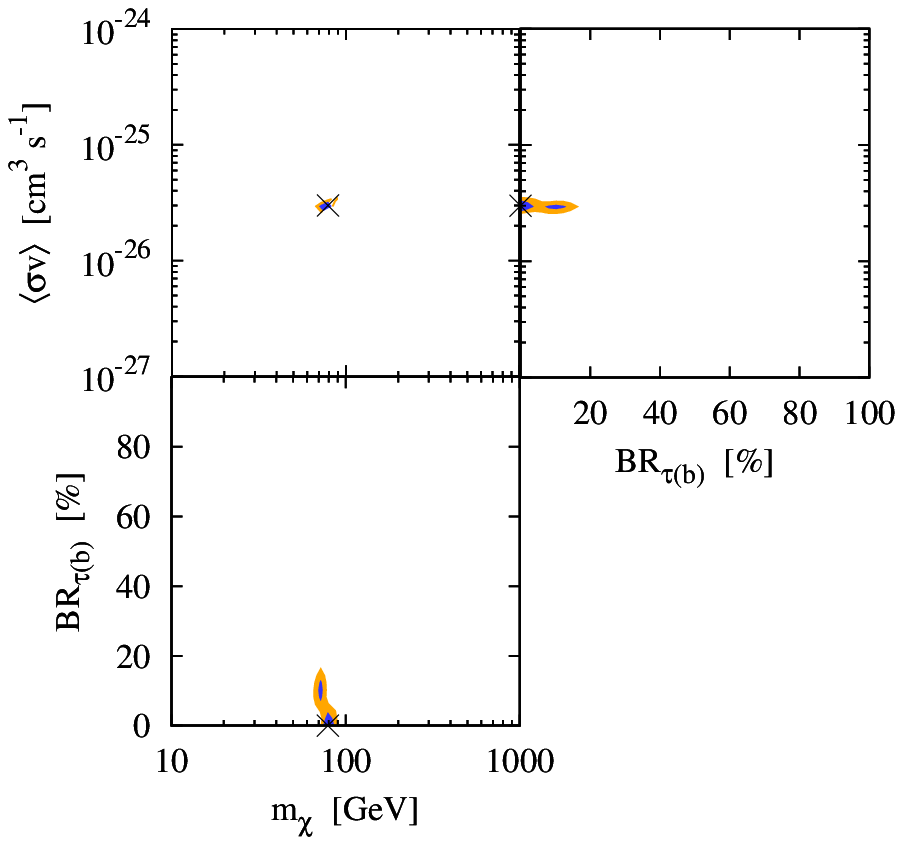}
\includegraphics[width=8.1cm]{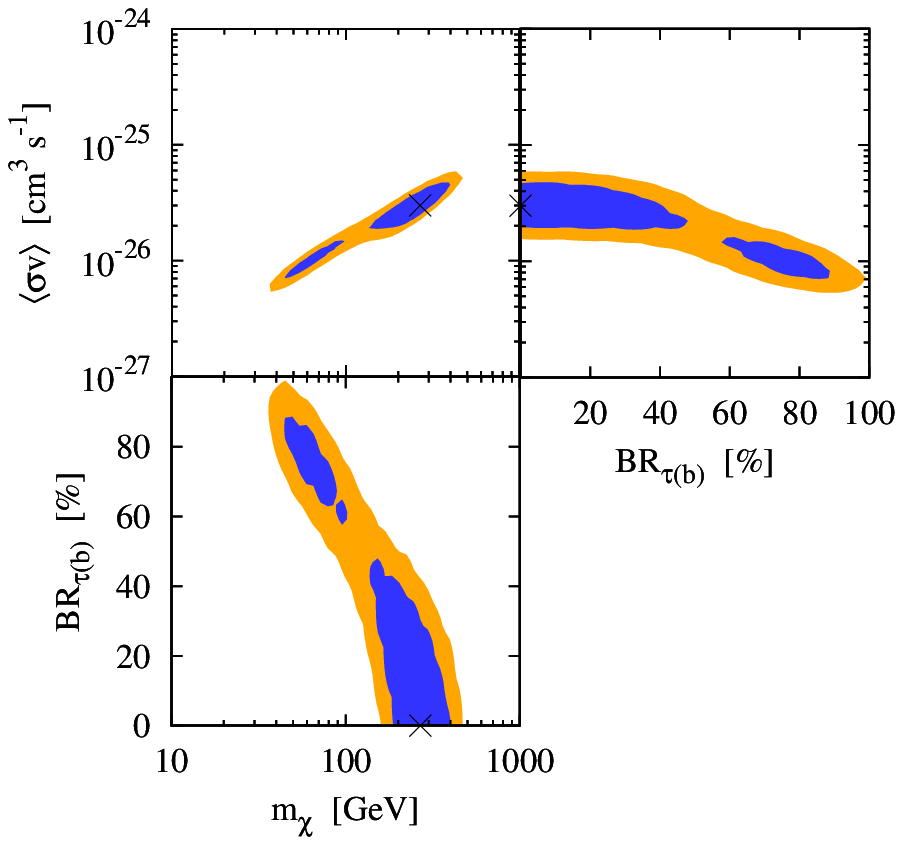}
\end{center}
\caption{\sl \textbf{\textit{Fermi--LAT abilities to constrain DM
  properties.}}  Same as Fig.~\ref{Fig:tau10deg} but for DM
  annihilations into a pure $b\bar b$ final state.}
\label{Fig:b10deg}
\end{figure}

The analysis in the previous section showed the region in the
parameter space for which DM could be distinguished from the
background in the GC for different cases of DM annihilations into pure
channels~\footnote{This is not meant to represent realistic examples
  of DM candidates, but just to allow for a model-independent approach.
  For a particular particle physics model, one would expect that
  annihilations would occur into a combination of different channels, so
  our results should be taken as limiting cases of realistic models.}.
Once this is accomplished, and gamma rays are identified as having
been produced in DM annihilations, the next step concerns the
possibilities of constraining DM properties.  Different approaches 
have been proposed to constrain DM properties by using indirect
searches, direct detection measurements, collider information or their
combination~\cite{Dodelson:2007gd, Bernal:2008zk, Bernal:2008cu,
  Jeltema:2008hf, PalomaresRuiz:2010pn, PalomaresRuiz:2010uu,
  Edsjo:1995zc, Cirelli:2005gh, Mena:2007ty, Agarwalla:2011yy,
  Das:2011yr, Lewin:1995rx, Primack:1988zm, Green:2007rb,
  Bertone:2007xj, Shan:2007vn, Drees:2008bv, Green:2008rd,
  Beltran:2008xg, Shan:2009ym, Strigari:2009zb, Peter:2009ak,
  Chou:2010qt, Shan:2010qv, Drees:2000he, Polesello:2004qy,
  Battaglia:2004mp, Allanach:2004xn, Weiglein:2004hn, Birkedal:2005jq,
  Moroi:2005zx, Nojiri:2005ph, Baltz:2006fm, Arnowitt:2007nt,
  Cho:2007qv, Arnowitt:2008bz, Belanger:2008yc, Cho:2008tj,
  Baer:2009bu, Bourjaily:2005ax, Altunkaynak:2008ry, Bertone:2010rv}.
These measurements are complementary and constitute an important step
toward identifying the particle nature of DM.  In this section we
discuss {\it Fermi}--LAT's abilities to constrain the DM mass,
annihilation cross section and the annihilation channels after 5~years
of data taking.  In principle, the analysis should include all
possible annihilation channels, but this would represent to have many
free parameters.  Nevertheless, in practice, they are commonly
classified as hadronic and leptonic channels.  Hence, for simplicity,
when simulating a signal, we will only consider two possible (generic)
channels.  This reduces the number of total free parameters to three:
the mass, $m_\chi$, the annihilation cross section, $\langle\sigma
v\rangle$, and the branching ratio into channel 1, $\text{BR}_{1(2)}$
(or equivalently into channel 2,
$\text{BR}_{2(1)}=1-\text{BR}_{1(2)}$).  We use the $\chi^2$ function
defined as
\begin{equation}
\chi^2  \left(m_\chi, \, \langle\sigma v\rangle, \, 
\text{BR}_{1(2)}\right) =  
\sum_{i=1}^{20}\frac{\left(S_i\left(m_\chi, \, \langle\sigma v\rangle,
  \, \text{BR}_{1(2)}\right)  - S_i^\text{th} \left(m_\chi^0, \,
  \langle\sigma v\rangle^0, \, \text{BR}_{1(2)}^0\right)
\right)^2}{S_i^\text{th} \left(m_\chi^0, \, \langle\sigma v\rangle^0,
  \, \text{BR}_{1(2)}^0 \right) +B_i}
~, 
\label{Eq:chi2results}
\end{equation}
where $S_i$ represents the simulated signal events in the $i$-th
energy bin for each set of the parameters and $S_i^\text{th}$ the
assumed observed signal events in that energy bin with parameters
given by $\left( m_\chi^0, \, \langle\sigma v\rangle^0, \,
\text{BR}_{1(2)}^0 \right)$.  

For our default setup we consider a $20^\circ \times 20^\circ$ (squared)
observational region around the GC, $\langle\sigma v\rangle^0=3\cdot
10^{-26}$~cm$^3$~s$^{-1}$, a NFW DM halo profile and the MED
propagation model.  Also by default, we consider annihilations into
$\tau^+\tau^-$ and $b\bar b$, both for the simulated observed ``data''
and the simulated signal.  However, we will also consider the case of
a simulated ``data'' with one channel and reconstructed signal with
other two different channels and add the $\mu^+\mu^-$ channel into the
analysis.  In Table~\ref{Tab:figs} we summarize the parameters used in
each of the figures we describe below.

\begin{figure}[t]
\begin{center}
\includegraphics[width=8.1cm]{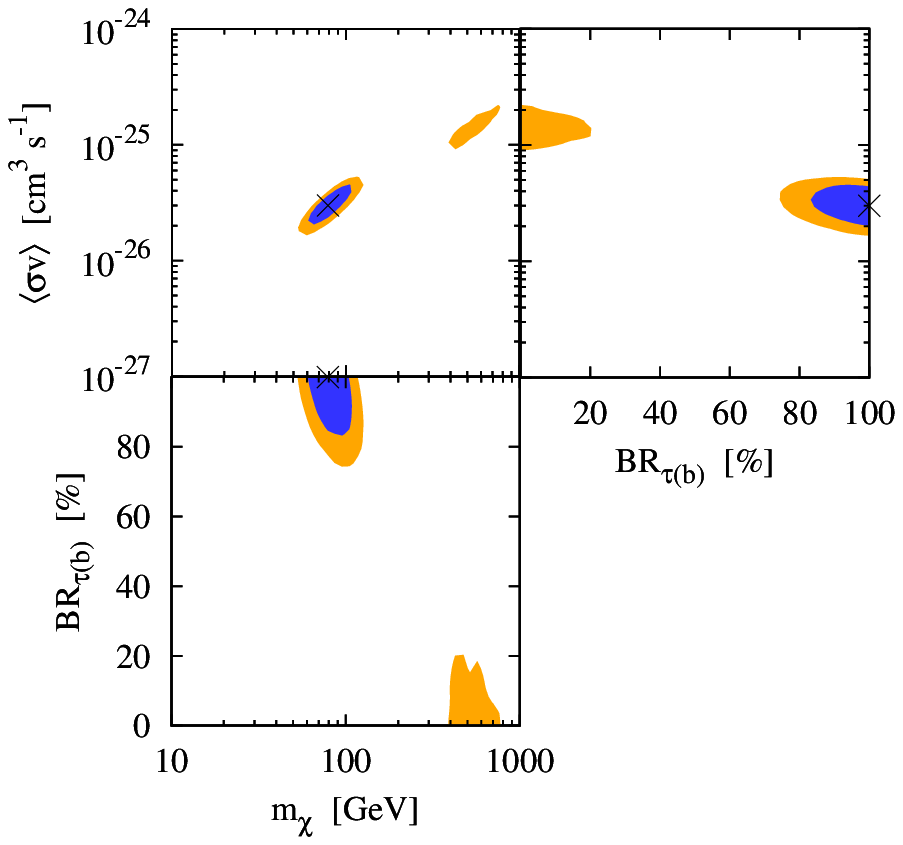}
\includegraphics[width=8.1cm]{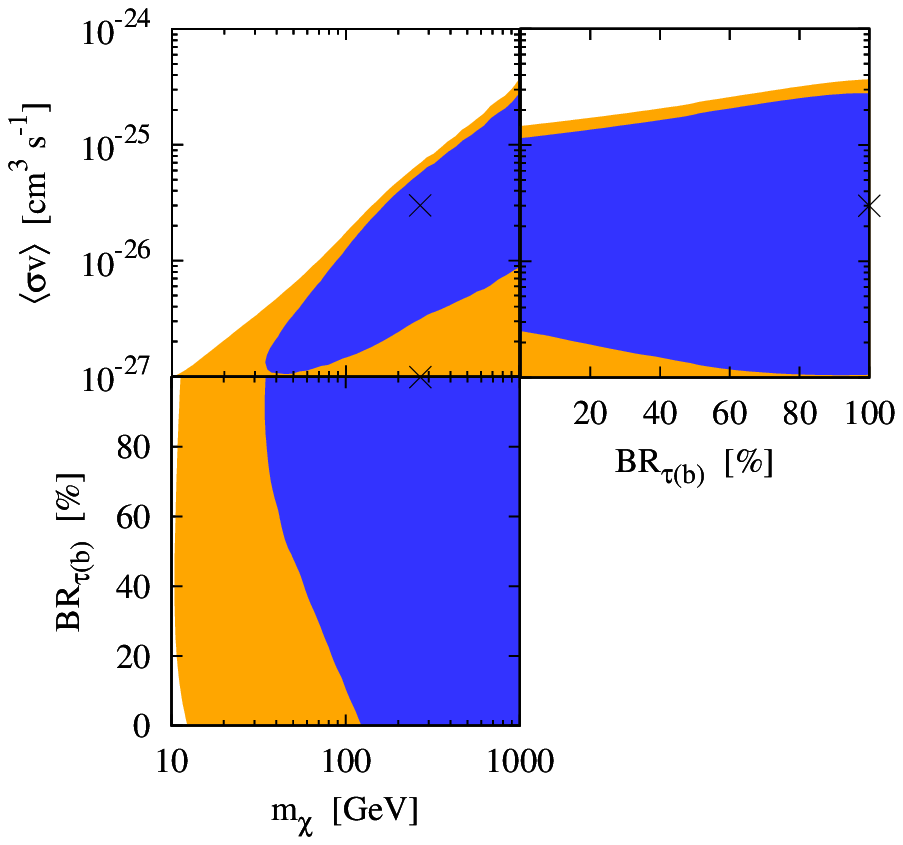}
\end{center}
\caption{\sl \textbf{\textit{Fermi--LAT abilities to constrain DM
  properties.}}  Same as Fig.~\ref{Fig:tau10deg} but for a $2^\circ
  \times 2^\circ$ observational region around the GC.}  
\label{Fig:tau1deg}
\end{figure}

\begin{figure}[t]
\begin{center}
\includegraphics[width=8.1cm]{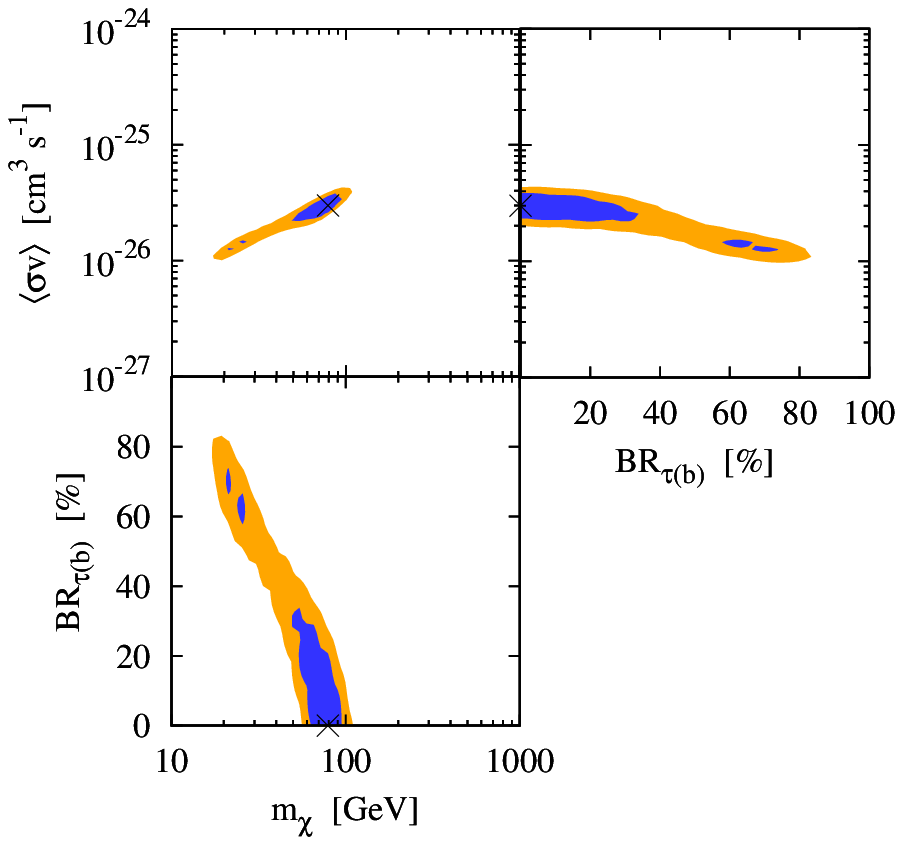}
\includegraphics[width=8.1cm]{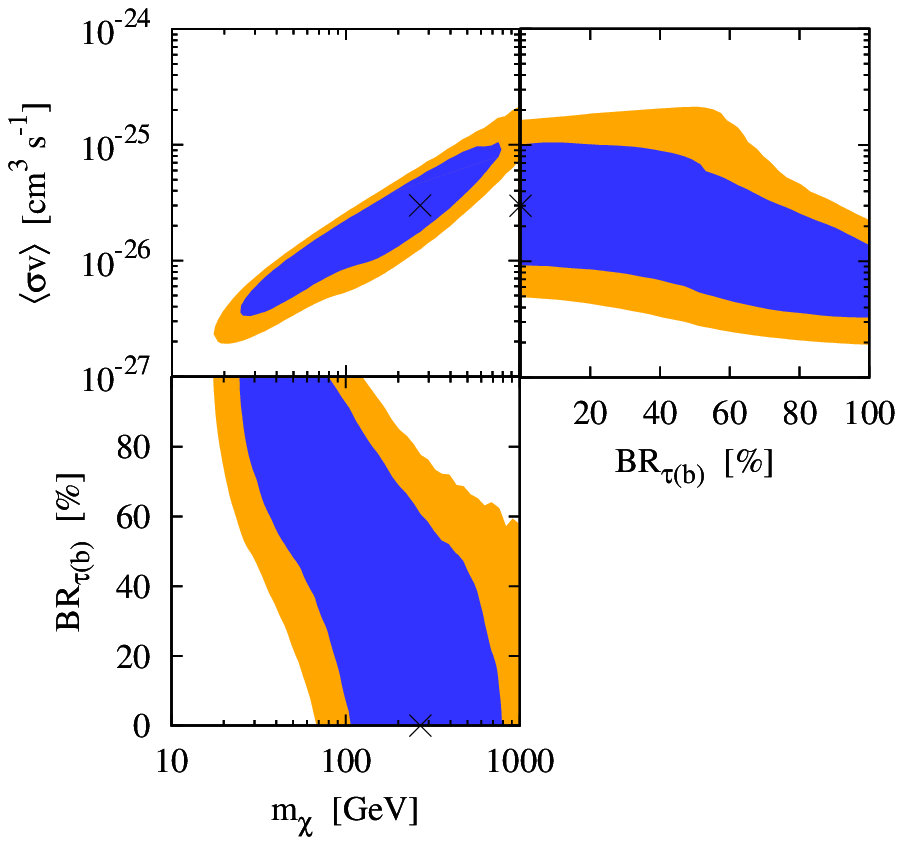}
\end{center}
\caption{\sl \textbf{\textit{Fermi--LAT abilities to constrain DM
  properties.}}  Same as Fig~\ref{Fig:b10deg} but for a $2^\circ \times
  2^\circ$ observational region around the GC.} 
\label{Fig:b1deg}
\end{figure}

In Fig.~\ref{Fig:tau10deg} we depict the {\it Fermi}--LAT reconstruction
prospects after 5~years for our default setup
(c.f.~Table~\ref{Tab:figs}), i.e., DM annihilation into a pure
$\tau^+\tau^-$ final state reconstructed as a combination of
$\tau^+\tau^-$ and $b\bar b$ and two possible DM masses:
$m_\chi^0=80$~GeV (left panels) and $m_\chi^0=270$~GeV (right panels).
By default, we also assume DM particle with an annihilation cross
section $\langle\sigma v\rangle=3\cdot 10^{-26}$~cm$^3$~s$^{-1}$, the
MED propagation model, a NFW DM halo profile and a $20^\circ \times
20^\circ$ observational region around the GC.  These benchmark points
are represented in the figure by black crosses.  The dark blue regions
and the light orange regions correspond to the $68\%$~CL and $90\%$~CL
contours (2 dof) respectively.  In Fig.~\ref{Fig:tau10deg}, the
different panels show the results for the planes
$(m_\chi,\,\langle\sigma v\rangle)$, $($BR$_{\tau(b)},\,\langle\sigma
v\rangle)$ and $(m_\chi,\,$BR$_{\tau(b)})$, marginalizing with respect
to the other parameter in each case.  For the first model chosen in
Fig.~\ref{Fig:tau10deg} (left panels), $m_\chi^0=80$ GeV, the
reconstruction prospects seem to be promising, allowing the
determination of the mass, the annihilation cross section and the 
annihilation channel at the level of $\sim$20\% or better.  For
lighter DM particles, the results substantially
improve~\cite{Jeltema:2008hf}.  Thus, for this cases, after 5~years of
data taking, {\it Fermi}--LAT could set very strong constraints on the
properties of DM.  On the other hand, for heavier DM particles, the
regions allowed by data grow considerably worsening the abilities of the
experiment to reconstruct DM properties.  This is shown for the second
model in Fig.~\ref{Fig:tau10deg} (right panels), $m_\chi^0=270$~GeV.  In
this case, {\it Fermi}--LAT would only be able to set a lower limit
(in the region of the parameter space we consider) on the DM mass
($m_\chi \gtrsim 130$~GeV at 90\%~CL for 2 dof) and to constrain the
annihilation cross section to be in the range $9 \cdot 10^{-27}
\lesssim \langle\sigma v\rangle \lesssim 2 \cdot
10^{-25}$~cm$^3$~s$^{-1}$ at 90\%~CL (2 dof).  Moreover, only at
68\%~CL (2 dof) some limited information about the annihilation
channel would be obtained.

On the other hand, Fig.~\ref{Fig:b10deg} depicts the {\it Fermi}--LAT
reconstruction abilities for a model similar to the one just
discussed, but assuming DM annihilates into a pure $b\bar b$ final
state, instead of $\tau^+\tau^-$.  As in the previous case, for light
DM particles (left panels) the reconstruction prospects are very good.
Note that the fact that the annihilation channel into $b\bar b$ has a
photon yield about an order of magnitude larger than the $\tau^+\tau^-$
channel does not necessarily imply that DM properties could be better
constrained in the $b\bar b$ case.  This can be understood by looking at
Fig.~\ref{Fig:signaltonoise}, where we can see that the
signal-to-noise ratio is similar for both cases.  This is due both to
the steep decrease of backgrounds with energy and to the fact that the
gamma-ray spectrum in the $\tau^+\tau^-$ case is harder and peaked
close to the DM mass, so fewer statistics are necessary to get a
reasonably good constrain on the DM mass.  As can be seen from
Fig.~\ref{Fig:b10deg}, in the $b\bar b$ case, for $m_\chi^0=270$~GeV
(right panels) and at 90\%~CL (2 dof), {\it Fermi}--LAT would be able to
constrain the DM mass to be in the range $\sim$(30--500)~GeV and
determine the annihilation cross section within an order of magnitude.

\subsection{Dependence on the observational region}

In Figs.~\ref{Fig:tau1deg} and~\ref{Fig:b1deg} we show the results
assuming the same properties for the DM particle as in
Figs.~\ref{Fig:tau10deg} and~\ref{Fig:b10deg}, respectively, but
assuming a $2^\circ \times 2^\circ$ observational region around the GC,
for which the background is dominated by resolved point sources.  As
can be seen from the figures,  the prospects of constraining DM
properties worsen in this case.  This was already expected from the
results in Refs.~\cite{Serpico:2008ga, Jeltema:2008hf}.  In this case,
second (spurious) minima appear at the 90\%~CL (2 dof) even for
$m_\chi = 80$~GeV at regions far from that of the simulated observed
``data''.  From Fig.~\ref{Fig:tau1deg} (left panels) we see that,
assuming $m_\chi^0 = 80$~GeV and annihilation into pure $\tau^+\tau^-$
final state, there is a small region at 90\%~CL (2 dof) reconstructed
with $m_\chi \simeq 690$~GeV and annihilation into pure $b\bar b$.
For $m_\chi^0 = 270$~GeV, basically no information would be extracted,
but just a very weak lower limit on the mass.  Similar results are
obtained for DM annihilation into $b\bar b$ as can be seen from
Fig.~\ref{Fig:b1deg}.  In this case, the second minimum for the
$m_\chi = 80$~GeV case appear at lower masses and annihilation cross
sections.  For larger masses very restricted information would be
available.  Like in Figs.~\ref{Fig:tau10deg} and~\ref{Fig:b10deg}, the
determination of the DM mass in this case is slightly worse than in
the case of annihilation into $\tau^+\tau^-$, even with better
statistics.

\subsection{Dependence on the DM density profile}

\begin{figure}[t]
\begin{center}
\includegraphics[width=8.1cm]{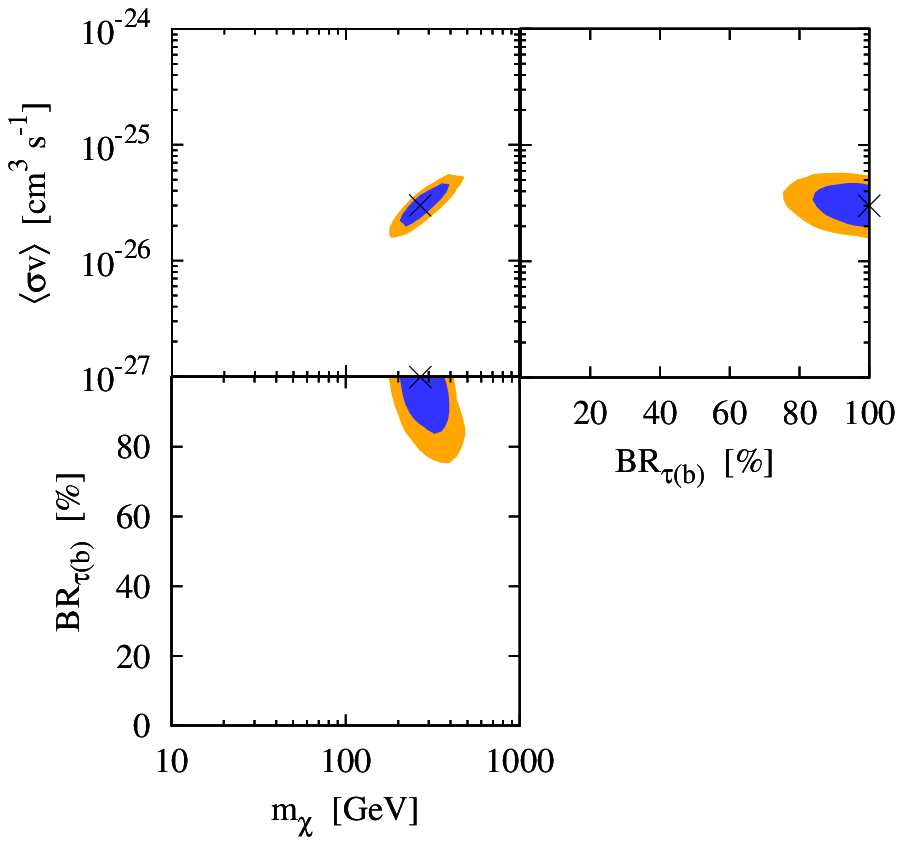}
\includegraphics[width=8.1cm]{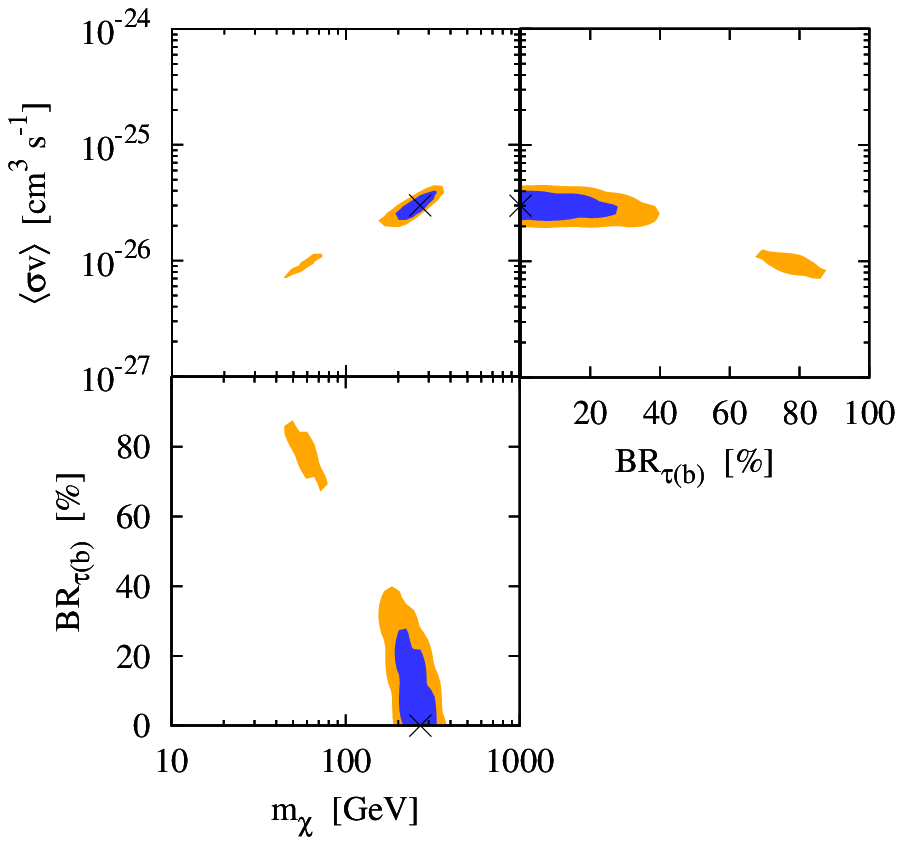}
\end{center}
\caption{\sl \textbf{\textit{Fermi--LAT abilities to reconstruct DM
  properties for a Einasto DM halo profile.}} We assume
  $m_\chi^0=270$~GeV and DM annihilation into $\tau^+\tau^-$ (left
  panels) and $b\bar b$ (right panels) final states. Dark blue (light
  orange) regions represent $68\%$~CL ($90\%$~CL) contours for 2 dof.
  See Table~\ref{Tab:figs} for the rest of the parameters.  The black
  crosses indicate the values of the parameters for the simulated
  observed ``data''.}
\label{Fig:einasto}
\end{figure}

On the other hand, recent state-of-the-art N-body numerical
simulations seem to converge towards a parameterization of the DM halo
profile described by the Einasto profile
(c.f. Eq.~\eqref{Eq:einasto})~\cite{Graham:2005xx, Graham:2006ae, 
  Graham:2006af, Navarro:2008kc}.  To illustrate this case, in
Fig.~\ref{Fig:einasto} we show how the results would improve if the
actual DM density profile is given by this parameterization.  In this
sense our previous results could be taken as a conservative approach.
We depict the results for $m_\chi^0=270$~GeV and for annihilations
into pure $\tau^+\tau^-$ (left panels) and $b\bar b$ (right panels) 
final states. As in the previous figures, we take a typical thermal
annihilation cross section, $\langle\sigma v\rangle^0=3\cdot
10^{-26}$~cm$^3$~s$^{-1}$, the MED propagation model and a
$20^\circ \times 20^\circ$ observational region around the GC.  The
left and right panels of this figure can be compared to the right
panels in Figs.~\ref{Fig:tau10deg} and~\ref{Fig:b10deg}, respectively.
Whereas in the case of a NFW profile only very limited information on
the DM mass could be obtained at 90\%~CL (2 dof), if DM is distributed
in the galaxy following a Einasto profile, the prospects would improve
substantially.  In the case of DM annihilation into a pure 
$\tau^+\tau^-$ channel (left panels), the DM mass could be constrained
with a $\sim$50\% uncertainty and the annihilation into $\tau^+\tau^-$
pairs with a branching ratio larger than 75\% established.  If DM
annihilates into $b \bar b$ pairs (right panels), except from a small
region present only at 90\%~CL (2 dof), DM mass could also be
determined with a $\sim$~50\% uncertainty and the annihilation
branching ratio into the right channel would be constrained to be
larger than 60\%.  As for the NFW case, in the Einasto case, for
lighter DM candidates the abilities of {\it Fermi}--LAT would improve,
whereas they would worsen if the DM particle is heavier.

Let us note that the profiles considered in this paper are obtained in
DM-only simulations that do not include baryons.  In principle, it is
not yet clear how baryons would affect the profile of the Milky Way,
but the picture could be significantly
changed~\cite{Blumenthal:1985qy, Prada:2004pi, Kazantzidis:2004vu,
  Gnedin:2004cx, Mambrini:2005vk, Gustafsson:2006gr, Pedrosa:2009bt,
  Abadi:2009ve, Tissera:2009cm, Duffy:2010hf, Kazantzidis:2010jp}.
One possibility is adiabatic contraction~\cite{Blumenthal:1985qy},
which would make the profiles steeper than in baryonless simulations,
and thus we would expect a larger signal that would improve the
detector sensitivity.  However, we leave the discussion of how our
results change due to the uncertainty on the DM density profile for
future work~\cite{Bernal:2011pz}.

\subsection{Dependence on systematic errors}

Now, let us discuss how our results are altered due to the
uncertainties in the gamma-ray background we are considering.  Here we
only show the effects for the case of the $20^\circ \times 20^\circ$
observational region around the GC, which is dominated by the DGE
below $\sim$~20~GeV.   As an illustration, we only consider the error
in the normalization of the total background, assigning a 20\%
uncertainty in its determination.  The treatment of this systematic
error is performed by the Lagrange multiplier method or also so-called
pull approach~\cite{Brock:2000ud, Pumplin:2000vx, Stump:2001gu,
  Fogli:2002pt}.  We use a nuisance systematic parameter that
describes the systematic error of the normalization of the background,
$\varepsilon_{\text{bkg}}$, and the variation of
$\varepsilon_{\text{bkg}}$ in the fit is constrained by adding a
quadratic penalty to the $\chi^2$ function, which in the case of a
Gaussian distributed error is given by
$\left(\varepsilon_{\text{bkg}}/\sigma_{\text{bkg}}\right)^2$, with 
$\sigma_{\text{bkg}}=0.2$ the standard deviation of the nuisance
parameter $\varepsilon_{\text{bkg}}$.  Hence, the simulated background 
events in each energy bin, $B_i$, are substituted by
$(1+\varepsilon_{\text{bkg}}) \, B_i$ and Eq.~\eqref{Eq:chi2results} 
is modified as  
\begin{equation}
\chi^2_{\text{pull}} = \text{min}_{\varepsilon_{\text{Bkg}}} \left\{  
\sum_{i=1}^{20}\frac{\left(S_i  + (1 + \varepsilon_{\text{bkg}}) \, B_i -
  S_i^\text{th} - B_i\right)^2}{S_i^\text{th} + B_i} +
\left(\frac{\varepsilon_{\text{bkg}}}{\sigma_{\text{bkg}}}\right)^2
\right\}~,
\label{Eq:chi2modresults}
\end{equation}
where $\chi^2_{\text{pull}}$ is obtained after minimization with
respect to the nuisance parameter $\varepsilon_{\text{bkg}}$.

\begin{figure}[t]
\begin{center}
\includegraphics[width=8.1cm]{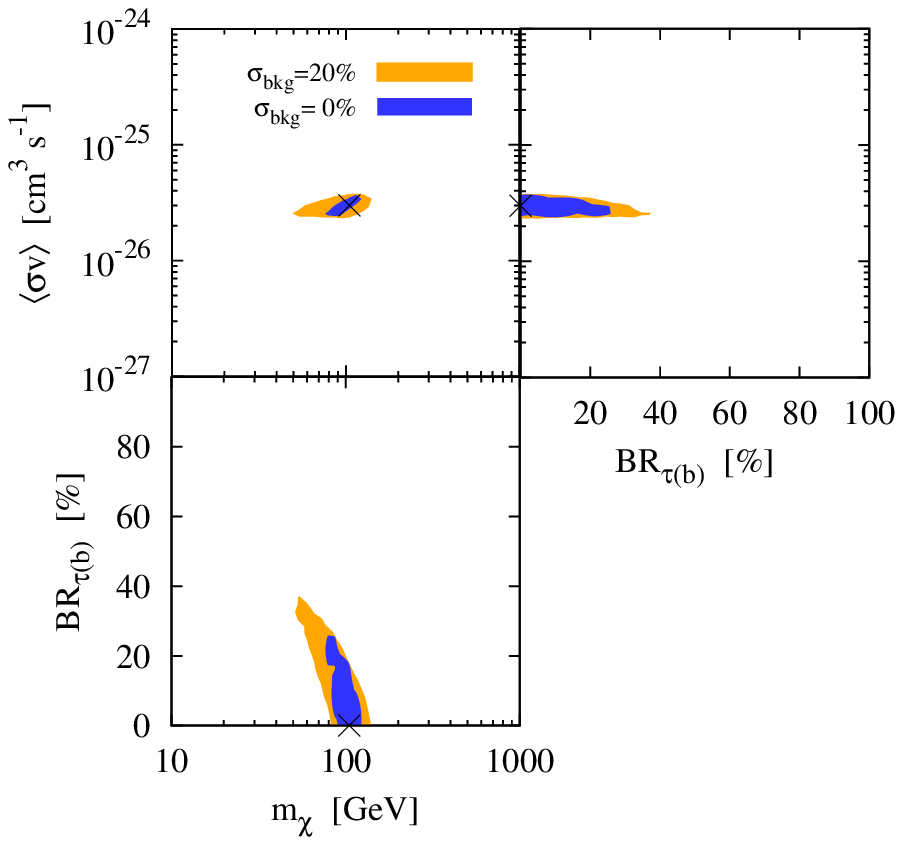}
\includegraphics[width=8.1cm]{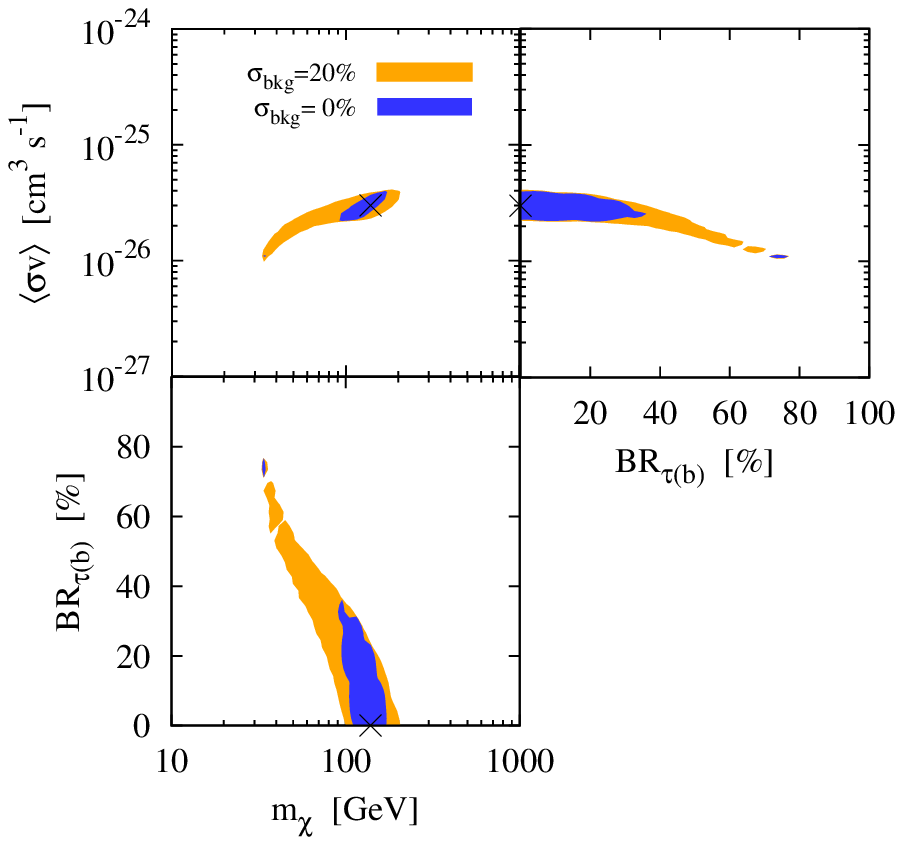}
\end{center}
\caption{\sl \textbf{\textit{Effects of adding a systematic error in
  the normalization of the gamma-ray background.}} We assume DM
  annihilation into pure $b\bar b$ final states and show the results
  at 90\%~CL (2 dof) for two DM masses: $m_\chi^0 = 105$~GeV (left
  panels) and $m_\chi^0 = 140$~GeV (right panels). Dark blue and light
  orange regions represent the case $\sigma_{\text{bkg}}=0$ (no error)
  and $\sigma_{\text{bkg}}=0.2$ in the gamma-ray background,
  respectively.  See Table~\ref{Tab:figs} for the rest of the
  parameters.  The black crosses indicate the values of the parameters
  for the simulated observed ``data''.}
\label{Fig:varyingbkg}
\end{figure}

The effects of adding this systematic error in the determination of the 
background are shown in Fig.~\ref{Fig:varyingbkg}.  We show the
results for DM annihilation into $b\bar b$ final states with 
$\langle\sigma v\rangle^0=3\cdot 10^{-26}$~cm$^3$~s$^{-1}$, assuming the
MED propagation model, a NFW DM halo profile and a $20^\circ \times
20^\circ$ observational region around the GC.  For $m_\chi^0 =
105$~GeV (left panels), the results do not change much, showing little
effect due to this uncertainty in the background.  However, for
$m_\chi^0 = 140$~GeV (right panels) the second spurious minimum
(c.f. right panels of Fig.~\ref{Fig:b10deg}) starts to show up when
adding the error in the background normalization, worsening the
results in a more significant way than for lighter masses.
Nevertheless, when the second minimum is already present in the case
of no error in the background (for slightly larger DM masses), taking
into account this error in the background has a negligible effect.
Thus, on general grounds, the error in the normalization of the
measured gamma-ray background we have studied here would not
substantially modify the results presented in this study regarding the
abilities of {\it Fermi}--LAT to constrain DM properties.  Let us
however note that the actual systematic uncertainties in the modeling
of the DGE are in principle larger than the considered 20\% error,
which could affect this analysis.

\begin{figure}[t]
\begin{center}
\includegraphics[width=8.1cm]{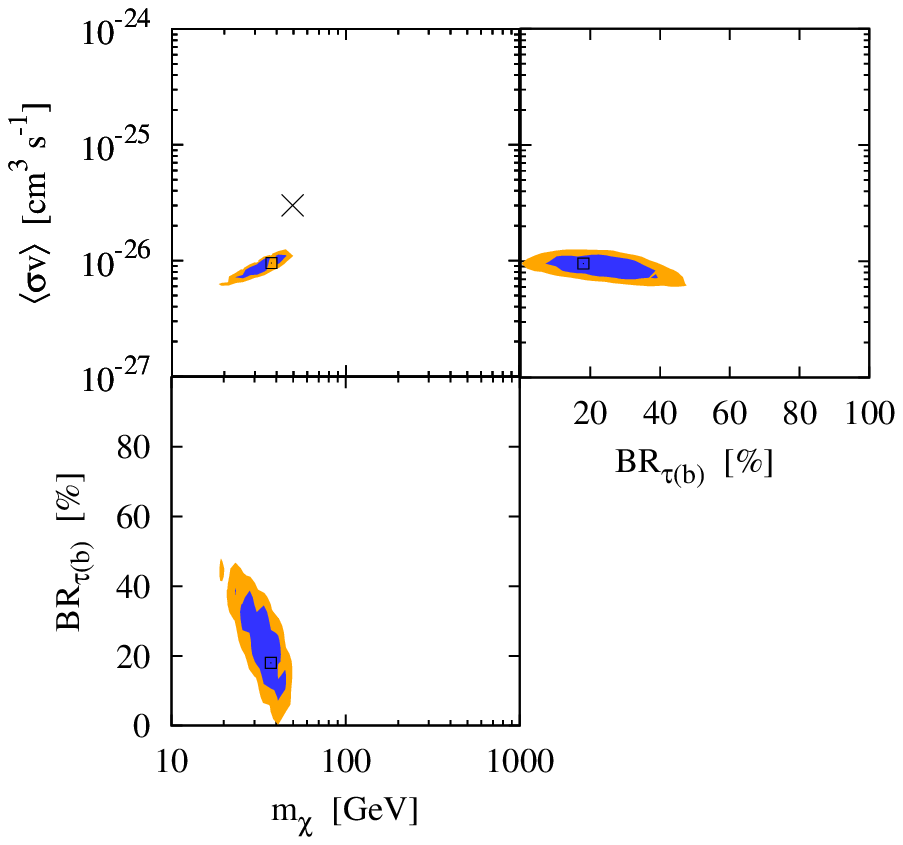}
\includegraphics[width=8.1cm]{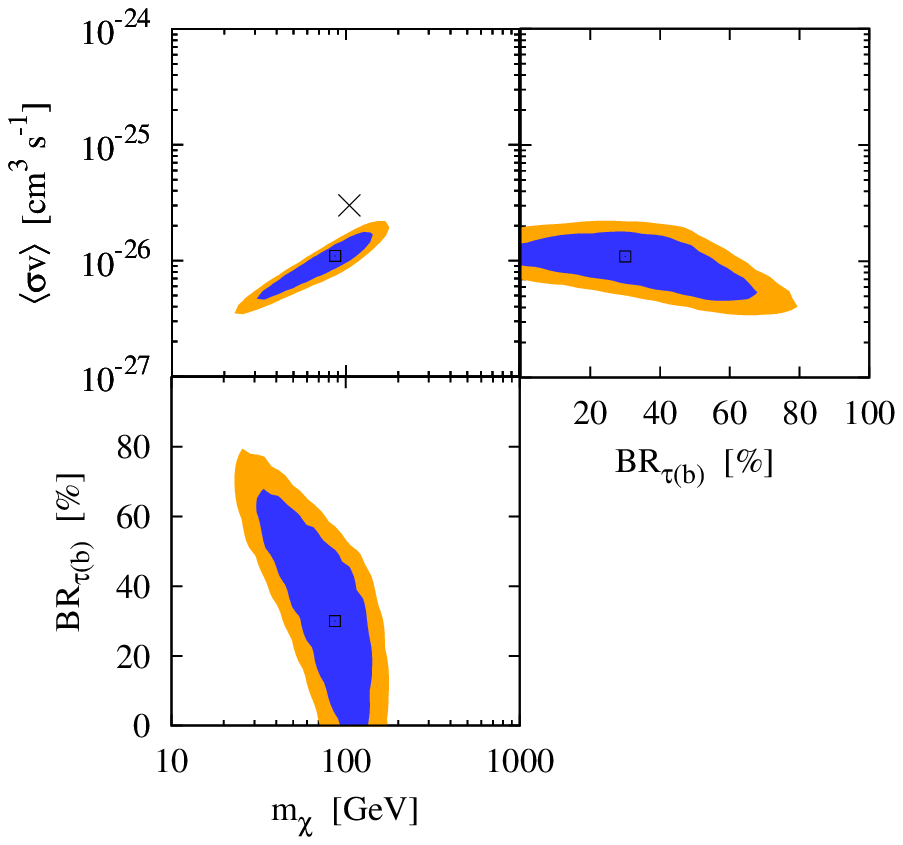}
\end{center}
\caption{\sl \textbf{\textit{Fermi--LAT abilities to constrain DM
  properties.}}  We assume the measured signal is due to DM
  annihilating into $\mu^+\mu^-$, but the fit is obtained assuming DM
  annihilates into a combination of $\tau^+\tau^-$ and $b\bar b$.  We
  assume two DM masses: $m_\chi^0=50$~GeV (left panels) and
  $m_\chi^0=105$~GeV (right panels).  Dark blue (light orange) regions
  represent the $68\%$~CL ($90\%$~CL) contours for 2 dof.  See
  Table~\ref{Tab:figs} for the rest of the parameters. The black cross
  in the left-top panel in each plot indicates the values of the
  parameters for the simulated observed ``data''.  Note that the other
  panels have no cross as they lie outside the parameter space of the
  simulated observed ``data''.  The squares indicate the best-fit
  point.}
\label{Fig:mubtau}
\end{figure}

\begin{figure}[t]
\begin{center}
\includegraphics[width=8.1cm]{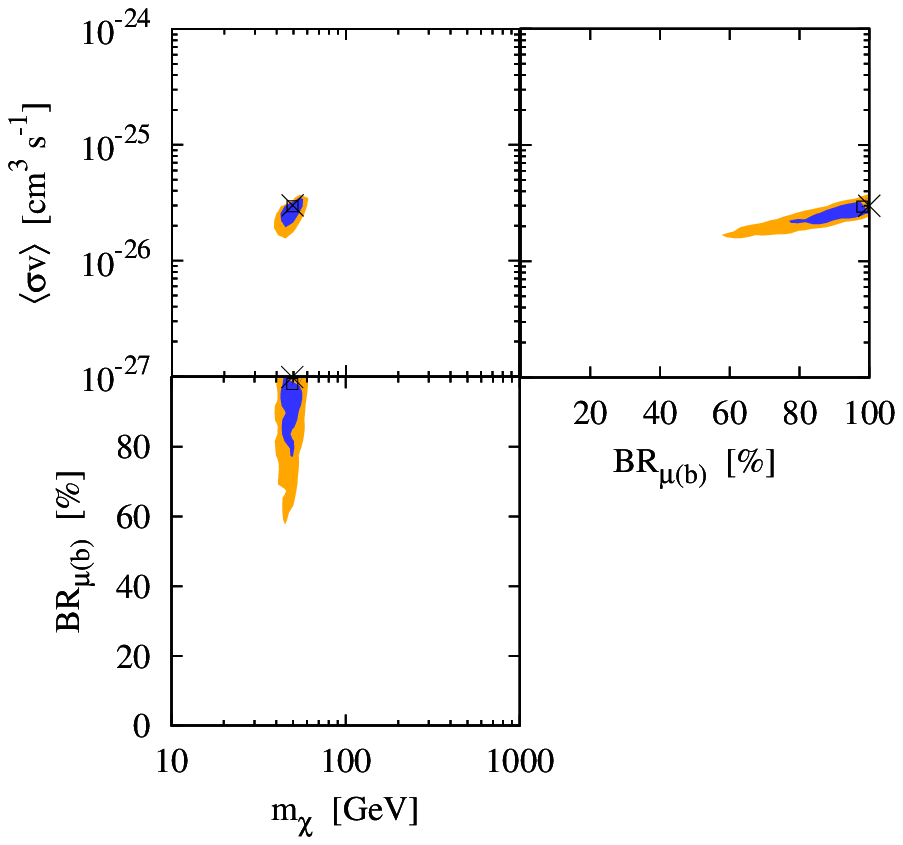}
\includegraphics[width=8.1cm]{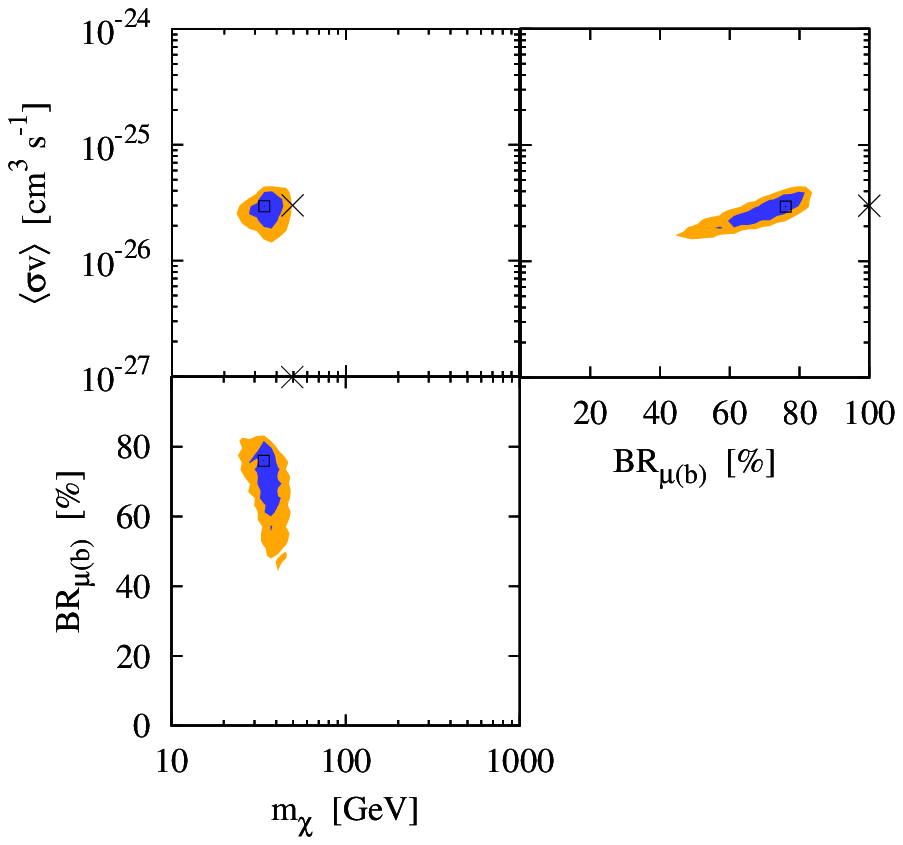}
\end{center}
\caption{\sl \textbf{\textit{Fermi--LAT abilities to constrain DM
  properties.}}  We assume the measured signal is due to DM
  annihilating into $\mu^+\mu^-$ and the fit is obtained assuming DM 
  annihilates into $\mu^+\mu^-$ or $b\bar b$.  We assume ICS+prompt
  photons (left panels) or only prompt photons (right panels) for the
  reconstructed signal, for $m_\chi^0=50$~GeV.  Dark blue (light
  orange) regions represent the $68\%$~CL ($90\%$~CL) contours for 2
  dof.  See Table~\ref{Tab:figs} for the rest of the parameters.  The
  black crosses indicate the values of the parameters for the simulated
  observed ``data''.  The squares in the right panels indicate the
  best-fit point.} 
\label{Fig:mumub}
\end{figure}

\subsection{Dependence on the assumed DM model}

Throughout this work we have so far considered that DM would either
annihilate into $\tau^+\tau^-$ or $b\bar b$ pairs or into a
combination of them.  This was in principle justified by the fact that
DM annihilation channels are commonly classified into two broad
classes: hadronic and leptonic channels.  However, we noted in
Section~\ref{sensitivity} that the contribution due to ICS in the case
of the $\mu^+\mu^-$ (and, not shown here, the $e^+e^-$) channel could
substantially alter the final sensitivity to DM annihilation from the
GC.  Thus, it is important to address the problem of assuming that DM
actually annihilates into $\mu^+\mu^-$ pairs, but we analyze the data
assuming DM annihilations into $\tau^+\tau^-$ and $b\bar b$.
Na\"{\i}vely, one would expect that the $\mu^+\mu^-$ (leptonic)
channel is identified as being closer to the $\tau^+\tau^-$ (leptonic)
channel than to the $b\bar b$ (hadronic) channel.  The results are
shown in Fig.~\ref{Fig:mubtau} for $\langle\sigma v\rangle^0=3\cdot
10^{-26}$~cm$^3$~s$^{-1}$, the MED propagation model, a NFW DM halo
profile, a $20^\circ \times 20^\circ$ observational region around the
GC and for two DM masses: $m_\chi^0 = 50$~GeV (left panels) and
$m_\chi^0 = 105$~GeV (right panels).  Contrary to what was expected,
the reconstructed composition of the annihilation channels tends to be
dominated by $b\bar b$, instead of $\tau^+\tau^-$.  Hence, when taking
into account the contribution of ICS to the gamma-ray spectrum, the
annihilation channels cannot be generically classified as hadronic or
leptonic, as DM annihilations into $\mu^+\mu^-$ pairs are better
reproduced with the $b\bar b$ channels than with the $\tau^+\tau^-$
channel.

The results just discussed can be illustrated in a different way by
analyzing the simulated observed signal ``data'' from DM annihilation
into $\mu^+\mu^-$ assuming DM annihilates into $\mu^+\mu^-$ and $b\bar
b$.  This is depicted in Fig.~\ref{Fig:mumub} where we show the
results for the case that we try to reconstruct the signal adding the
ICS contribution (left panels) or with only prompt photons (right
panels). As can be seen in the left panels, if ICS is taken into 
account, DM properties can be reconstructed with good precision.
However, if the ICS contribution is not added to the simulated signal
events (the simulated observed ``data'' always has the ICS included),
DM annihilation into a pure $\mu^+\mu^-$ channel would be excluded at
more 90\%~CL (2 dof), providing thus a completely wrong result.

\begin{figure}[t]
\begin{center}
\includegraphics[width=8.1cm]{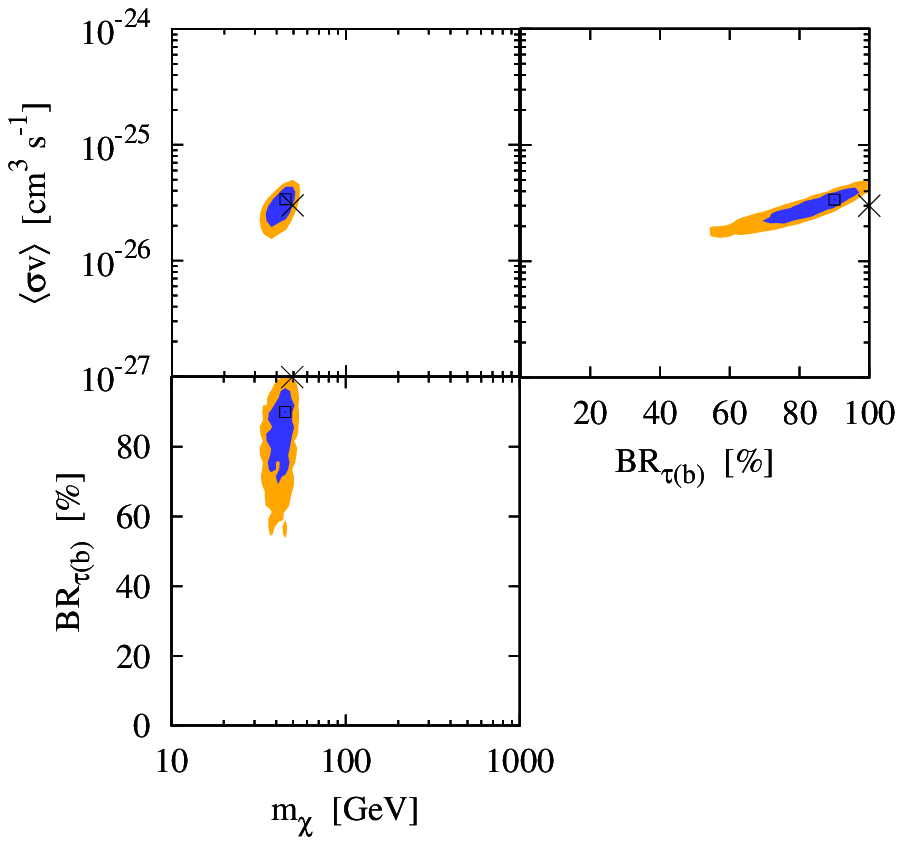}
\includegraphics[width=8.1cm]{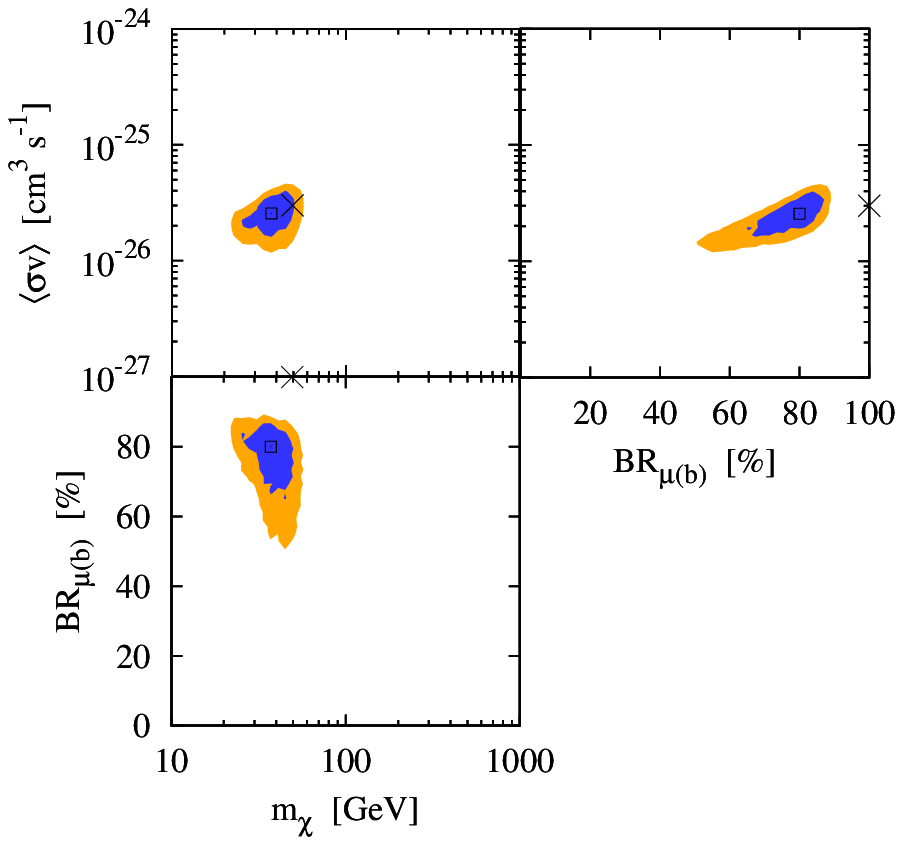}
\end{center}
\caption{\sl \textbf{\textit{Fermi--LAT abilities to constrain DM
  properties.}}  Same as Fig.~\ref{Fig:mumub}, but assuming the MAX
  propagation model for the simulated observed ``data'' and the MIN
  propagation model for the simulated signal.}
\label{Fig:prop}
\end{figure}

We conclude that, although the inclusion of the ICS contribution for
hadronic channels and for the $\tau^+\tau^-$ channel does not give
rise to important differences in the presented results, this is not
the case if DM annihilates into the $\mu^+\mu^-$ (and, not shown here,
the $e^+e^-$) channel.  In this latter case, adding the ICS
contribution to the prompt gamma-ray spectrum renders crucial in order
not to obtain completely wrong results.

\subsection{Dependence on the propagation model}

We have just seen that taking into account the contribution from ICS
turns out to be fundamental if the DM signal is produced from DM
annihilations into $\mu^+\mu^-$ (and $e^+e^-$).  However, there are
still a number of different uncertainties on the propagation of
electrons and positrons in the Galaxy, which directly affect on the
final ICS contribution to the DM-induced gamma-ray spectrum.  Thus,
the natural question to address is determining what the effect of
these uncertainties is on the results presented here.  In
Fig.~\ref{Fig:prop} we repeat Fig.~\ref{Fig:mumub}, but in this case
we take the MAX propagation model for the simulated observed ``data''
and the MIN propagation model for the simulated signal.  Likewise, we
constrain the signal adding the ICS contribution (left panels) or with
only prompt photons (right panels).  As can be seen by comparing both
figures, the effect of adding the electron and positron propagation
uncertainties does not alter significantly the conclusions reached with
Fig.~\ref{Fig:mumub}.  Most importantly, if the ICS component is not
added to the simulated signal, DM annihilation only into the
$\mu^+\mu^-$ channel would also be excluded at much more than 90\%~CL
(2 dof), and thus leading to misidentification of the nature of DM.   

Finally, let us note that in this work we are using a simplified
position-independent parameterization of the ISRF.  A more detailed
modelization would be required to compute more accurately the final
ICS gamma-ray contribution to the potential signal and to fully take
into account the effect of the induced uncertainties.  However, the
anisotropy of the background radiation field is not expected to induce 
corrections larger than ${\cal O} (15\%)$ to the ICS contribution
around the GC~\cite{Moskalenko:1998gw}.  Another
source of uncertainty comes from the fact that $e^+e^-$ produced
outside the diffusive zone can enter and get trapped in it and thus
give rise to an increased ICS gamma-ray flux by ${\cal O}
(20\%)$~\cite{Perelstein:2010fq}.

\section{Conclusions}
\label{conclusions}

One of the most important topics in current astroparticle physics is
the physics of DM.  Its discovery and the determination of its nature
once it is detected play a central role in astrophysics, cosmology and
particle physics research.  This is specially exciting in the light of
recent data from direct detection experiments, which have further
improved existing bounds and even hinted for a possible DM
signal~\cite{Bernabei:2010mq, Aalseth:2010vx, Aalseth:2011wp,
  Angloher:2011uu}.  In addition to direct searches, there are other
approaches that have been considered to detect DM: collider
experiments could find evidences for the presence of particles beyond
the SM which could be good DM candidates and indirect searches looking
for the products of DM annihilation (or decay), as antimatter,
neutrinos and photons.  Indeed, also hints of a DM signal have been
recently suggested in indirect detection
experiments~\cite{Adriani:2008zr, Adriani:2010ib, :2008zzr,
  Torii:2008xu, Abdo:2009zk, Grasso:2009ma, Aharonian:2009ah}.

Once DM has been detected and identified, the next step would be to
use the available information from different experiments and constrain
its properties.  Different approaches have been proposed to determine
DM properties by using future indirect DM-induced signals of
gamma-rays or neutrinos, direct detection measurements, collider
information or their combination~\cite{Dodelson:2007gd, Bernal:2008zk,
  Bernal:2008cu, Jeltema:2008hf, PalomaresRuiz:2010pn,
  PalomaresRuiz:2010uu, Edsjo:1995zc, Cirelli:2005gh, Mena:2007ty,
  Agarwalla:2011yy, Das:2011yr, Lewin:1995rx, Primack:1988zm,
  Green:2007rb, Bertone:2007xj, Shan:2007vn, Drees:2008bv,
  Green:2008rd, Beltran:2008xg, Shan:2009ym, Strigari:2009zb,
  Peter:2009ak, Chou:2010qt, Shan:2010qv, Drees:2000he,
  Polesello:2004qy, Battaglia:2004mp, Allanach:2004xn,
  Weiglein:2004hn, Birkedal:2005jq, Moroi:2005zx, Nojiri:2005ph,
  Baltz:2006fm, Arnowitt:2007nt, Cho:2007qv, Arnowitt:2008bz,
  Belanger:2008yc, Cho:2008tj, Baer:2009bu, Bourjaily:2005ax,
  Altunkaynak:2008ry, Bertone:2010rv}.

In this work we have studied the abilities of the {\it Fermi}--LAT
instrument on board of the {\it Fermi} mission to constrain DM
properties by using the current and future observations of gamma-rays
from the Galactic Center (GC) produced by DM annihilations.  Unlike
previous works~\cite{Dodelson:2007gd, Bernal:2008zk, Bernal:2008cu,
  Jeltema:2008hf}, we also take into account the contribution to the
gamma-ray spectrum from ICS of electrons and positrons produced in DM
annihilations off the ambient photon background, which in the case of
DM annihilations into $\mu^+\mu^-$ and $e^+e^-$ pairs turns out to be
crucial.

After an introductory review of the main components of the gamma-ray
emission from DM annihilations in the GC, where we explicitly write
the relevant formulae, we discuss the three main gamma-ray foregrounds
in the GC, which we obtain after performing an unbinned likelihood
analysis (with the {\it Fermi} Science Tools) of the {\it Fermi}--LAT
data collected from August 4, 2008 to July 12, 2011.  We notice that
for a $2^\circ \times 2^\circ$ observational region around the GC, the
high concentration of resolved point sources in the GC provides the
dominant gamma-ray background above $\sim$~10~GeV.  However, for
larger observational regions, as that with a field of view of
$20^\circ \times 20^\circ$ around the GC, the density of point sources
dilutes and the dominant background is the DGE, primarily originated
from the interaction of cosmic rays with the interstellar nuclei and
the ISRF.

In Section~\ref{sensitivity} we describe the {\it Fermi}--LAT
instrument on board of the {\it Fermi} mission, which we simulate by
means of the tools provided by the collaborations.  We first study the
signal-to-noise ratio, Fig.~\ref{Fig:signaltonoise}, to understand the
relevance of the different components of the signal.  We see that,
although in principle the inclusion of the ICS contribution is
important for all leptonic channels, this does not seem to be the case
for DM annihilations into the $\tau^+\tau^-$ channel and in the energy
range under study (1--300~GeV).  Nevertheless, the ICS contribution is
the most important one for the $\mu^+\mu^-$ (and, not discussed here,
the $e^+e^-$) channel.

We then evaluate the {\it Fermi}--LAT sensitivity to DM annihilation
in the GC after 5~years of data taking by observing a $20^\circ \times
20^\circ$ region around the GC.  We show the results in
Fig.~\ref{Fig:exclu} for two different DM halo profiles, NFW and
Einasto, and study the effect of the uncertainties in the propagation 
parameters, which turns out to be small in this observational region.
We note that had we considered a smaller region of observation, these
uncertainties would have had a more important effect.  We see that for
DM candidates lighter than 1~TeV annihilating into two SM particles,
an overall conservative bound at 90\%~CL (1 dof) of $\langle\sigma
v\rangle < 10^{-25}$~cm$^3$~s$^{-1}$ is obtained.  The bound improves
by more than two orders of magnitude for lighter DM particles and for
annihilations into hadronic channels, being always below the benchmark
value for thermal DM, $\langle\sigma v\rangle^0=3\cdot
10^{-26}$~cm$^3$~s$^{-1}$, for all masses below 1~TeV and
annihilations into $b\bar b$ (or generically into hadronic channels). 

In Section~\ref{results} we present the core results of our paper.  We
describe in detail the {\it Fermi}--LAT prospects for constraining DM
properties for different DM scenarios.  We also show the dependences
on several assumptions and how they may affect the abilities of the
experiment to determine some DM properties, as DM annihilation cross
section (times relative velocity), DM mass and branching ratio into
dominant annihilation channels.  Throughout the analysis, we have
considered a default setup defined by a $20^\circ \times 20^\circ$
observational region around the GC, $\langle\sigma v\rangle^0=3\cdot
10^{-26}$~cm$^3$~s$^{-1}$, a NFW DM halo profile and the MED
propagation model for electrons and positrons.  Also by default, we 
have considered annihilations into $\tau^+\tau^-$ and $b\bar b$, both
for the simulated observed ``data'' and the simulated signal.  We
provide Table~\ref{Tab:figs} where we summarize the different
parameters used in each of the figures described in
Section.~\ref{results}.

Along the various figures in that section, we show the {\it Fermi}--LAT
abilities to determine DM properties after 5~years of data taking, which
is our default observation time.  In Fig.~\ref{Fig:tau10deg} we show
the reconstruction prospects for our default setup (DM annihilations
into $\tau^+\tau^-$).  In  Fig.~\ref{Fig:b10deg} we just change the
simulated observed ``data'' and assume that DM annihilates into $b\bar
b$. With these figures we show the general trend of our results.  For
DM masses below $\sim$~200~GeV, {\it Fermi}--LAT will likely be able
to constrain the annihilation cross section, DM mass and dominant
annihilation channels with good accuracy.  The larger the DM mass, the
more difficult this task will be, as the signal decreases with the DM
mass.

Finally, we also study the effects of the dependence on different
parameters, like the region of observation (Figs.~\ref{Fig:tau1deg}
and~\ref{Fig:b1deg}), the DM density profile (Fig.~\ref{Fig:einasto}),
the particular assumptions for the DM model concerning annihilation
channels (Figs.~\ref{Fig:mubtau} and~\ref{Fig:mumub}) and the
uncertainties in the propagation model (Fig.~\ref{Fig:prop}).  We
also show the effect of the inclusion of a 20\% systematic uncertainty
in the gamma-ray background (Fig.~\ref{Fig:varyingbkg}).  As mentioned
above, one important result is that the commonly used classification
of annihilation channels into hadronic or leptonic, based on the
prompt gamma-ray spectrum, is not so clear when the ICS contribution
is added.  Indeed, DM annihilation into $\mu^+\mu^-$ could be better
fitted by DM annihilation into $b\bar b$ than into $\tau^+\tau^-$.  In
addition, if the ICS contribution is not included in the simulated
signals, one could reach wrong conclusions, as for instance
incorrectly excluding DM annihilations into $\mu^+\mu^-$ when the data
is actually due to this annihilation channel.  Hence, this is
particularly important for any realistic model with significant dark
matter annihilations into $\mu^+\mu^-$ or $e^+e^-$, as for instance,
any leptophilic DM models.

\begin{table}
\begin{center}
\begin{tabular}{|c|c|c|c|c|c|c|c|c|}
\cline{2-9}
\multicolumn{1}{c|}{}
  & $\theta_\text{max}$  & ``Data'' & Signal & $m_\chi^0$ & DM Profile &
$\sigma_{\text{bkg}}$ & Prop. Model & $\langle\sigma v\rangle^0$ \\  
\multicolumn{1}{c|}{} & & & & [GeV] & & & & [cm$^3$~s$^{-1}$] \\
\hline
{\bf DS} &  {$\boldsymbol{20^\circ}$} & {$\boldsymbol{\tau^+\tau^-}$} &
{$\boldsymbol{\tau^+\tau^-/b\bar b}$} & {\bf 80/270} & {\bf NFW} &
{\bf 0} & {\bf MED} & {$\boldsymbol{3 \cdot 10^{-26}}$} \\ 
\hline 
Fig.~\ref{Fig:tau10deg} 
& - & - & - & - & - & - & - & - \\[1ex] 
Fig.~\ref{Fig:b10deg}  
& - & $b\bar b$ & - & - & - & - & - & - \\[1ex] 
Fig.~\ref{Fig:tau1deg} 
& $2^\circ$ & - & - & - & - & - & - & - \\[1ex] 
Fig.~\ref{Fig:b1deg}
& $2^\circ$ & $b\bar b$ & - & - & - & - & - & - \\[1ex]  
Fig.~\ref{Fig:einasto}
& - & $\tau^+\tau^-$/$b\bar b$ & - & 270 & Einasto & - & - & - \\[1ex] 
Fig.~\ref{Fig:varyingbkg} 
& - & $b\bar b$ & - & 105/140 & - & 0.2 & - & - \\[1ex] 
Fig.~\ref{Fig:mubtau}  
& - & $\mu^+\mu^-$ & - & 50/105 & - & - & - & - \\[1ex]   
Fig.~\ref{Fig:mumub}
& - & $\mu^+\mu^-$ & $\mu^+\mu^-$/$b\bar b$ & 50 & - & - & - & -
\\[1ex]   
Fig.~\ref{Fig:prop}
& - & $\mu^+\mu^-$ & $\mu^+\mu^-$/$b\bar b$ & 50 & - & - & MAX/MIN & -
\\ 
\hline
\end{tabular}
\caption{\sl \textbf{\textit{Summary of the parameters used in each of
  the figures in Section~\ref{results}.}}  The default setup is
  referred to as DS and $\theta_{\text{max}}$ indicates the size of
  the observational $\theta_{\text{max}} \times \theta_{\text{max}}$
  region around the GC.  We indicate by `-' when the parameters are
  the default ones.  All the figures assume 5~years of data taking.
} 
\label{Tab:figs}
\end{center}
\end{table}

All in all, we would like to stress that the first task for any
experiment aiming to detect DM is to distinguish it from any other
possible source of signal or background.  Here we have considered the
GC, one of the most complex regions in the sky, which turns the
modeling of the gamma-ray background into a difficult problem.  In
this context, multiwavelength studies will definitely be crucial to
reduce all related uncertainties and reach a satisfactory
understanding of the galactic foregrounds~\cite{Regis:2008ij,
  Crocker:2010gy, Profumo:2010ya}.  Baring carefully in mind these
issues, our study shows the {\it Fermi}--LAT capabilities to constrain
DM properties and can be used as a starting point for more detailed
analysis, which will be needed when a convincing signal is detected,
hopefully in the near future.

\section*{Acknowledgments}
We thank G.~A.~Gómez for helpful information about the {\it Fermi}
Science Tools and data.  We also thank the Galileo Galilei Institute,
where this work was finished, for hospitality.  NB is supported by the
EU project MRTN-CT-2006-035505 HEPTools and the DFG TRR33 `The Dark
Universe'.  SPR is partially supported by the Portuguese FCT through
CERN/FP/109305/2009, CERN/FP/116328/2010 and CFTP-FCT UNIT 777, which
are partially funded through POCTI (FEDER), and by the Spanish Grant
FPA2008-02878 of the MICINN.


\small
\bibliographystyle{utphys}
\addcontentsline{toc}{section}{References}
\bibliography{biblio}

\end{document}